\documentclass[journal]{IEEEtran}
\usepackage[cmex10]{amsmath}
\usepackage{graphicx,amssymb,amsthm,comment,bm,cite,url}
\usepackage{color}
\usepackage{boldline}
\usepackage{algorithm,algorithmic}
\usepackage{multirow}
\usepackage {colortbl,array,xcolor}
\usepackage{booktabs}
\allowdisplaybreaks

\definecolor{color1}{rgb}{0.957,0.8,0.8}
\definecolor{color2}{rgb}{0.988,0.898,0.804}
\definecolor{color3}{rgb}{1.0,0.949,0.8}
\definecolor{color4}{rgb}{0.851,0.918,0.827}
\definecolor{color5}{rgb}{1,1,1}

\newcommand{\refeq}[1]{Eq. (\ref{eq:#1})}

\newcommand{\refsubsec}[1]{Subsection \ref{subsec:#1}}

\newcommand{\reffig}[1]{Fig. \ref{fig:#1}}

\newcommand{\reffigss}[2]{Figs. \ref{fig:#1}--\ref{fig:#2}}
\newcommand{\reftab}[1]{Table \ref{tab:#1}}
\newcommand{\reftabs}[2]{Tables \ref{tab:#1} and \ref{tab:#2}}

\newcommand{\refalgo}[1]{Algorithm \ref{algo:#1}}
\newcommand{\refalgos}[2]{Algorithms \ref{algo:#1} and \ref{algo:#2}}

\def\Vec#1{\boldsymbol{\mathbf{#1}}}

\def\thline{\noalign{\hrule height 1.2pt}}

\ifCLASSINFOpdf
\else
\fi

\begin{document}

\title{
LatentVoiceGrad: 
Nonparallel Voice Conversion with Latent Diffusion/Flow-Matching Models
}

\author{Hirokazu~Kameoka,~
Takuhiro Kaneko,
Kou Tanaka,
and 
Yuto Kondo%
\thanks{H. Kameoka, T. Kaneko, K. Tanaka, 
and Y. Kondo are with NTT Communication Science Laboratories, NTT, Inc., Atsugi, Kanagawa, 243-0198 Japan (e-mail: hirokazu.kameoka@ntt.com).}
}

%



\maketitle

\begin{abstract}
Previously, we introduced VoiceGrad, a nonparallel voice conversion (VC) technique enabling mel-spectrogram conversion from source to target speakers using a score-based diffusion model. The concept involves training a score network to predict the gradient of the log density of mel-spectrograms from various speakers. VC is executed by iteratively adjusting an input mel-spectrogram until resembling the target speaker's. However, challenges persist: audio quality needs improvement, and conversion is slower compared to modern VC methods designed to operate at very high speeds. To address these, we introduce latent diffusion models into VoiceGrad, proposing an improved version with reverse diffusion in the autoencoder bottleneck. 
Additionally, we propose using a flow matching model as an alternative to the diffusion model to further speed up the conversion process without compromising the conversion quality.
Experimental results show enhanced speech quality and accelerated conversion compared to the original.
\end{abstract}


%
\IEEEpeerreviewmaketitle

\section{Introduction}
\label{sec:intro}

Voice conversion (VC) is a technique used to alter the voice of one speaker to resemble that of another without altering the content of the spoken sentence. Its uses vary widely, encompassing personalized speech synthesis \cite{Kain1998short}, speech assistance \cite{Kain2007short,Nakamura2012short}, speech enhancement \cite{Inanoglu2009short,Turk2010short,Toda2012short}, bandwidth extension \cite{Jax2003short}, accent conversion \cite{Felps2009short}, and privacy protection \cite{Srivastava2020}.

Many traditional VC techniques rely on parallel utterances to train acoustic models for feature mapping. In contrast, non-parallel VC methods are capable of functioning without requiring parallel utterances or transcriptions during model training. These methods offer significant advantages as constructing a parallel corpus is often prohibitively expensive and lacks scalability. Additionally, achieving any-to-any (or any-to-many) conversion is another crucial requirement for VC methods, allowing them to convert the speech of any speaker to the voices of speakers not originally included in the training dataset.
Such methods are appealing because they can process input speech from unknown speakers without necessitating model retraining or adaptation.

Several non-parallel methods have been put forward, with notable attention in recent years focused on those grounded in deep generative models. These include variational autoencoders (VAEs) \cite{Kingma2014ashort,Kingma2014bshort}, generative adversarial networks (GANs) \cite{Goodfellow2014short}, flow-based models \cite{Dinh2015,Dinh2017,Kingma2018}, and score-based generative models \cite{Song2019,Song2020,Ho2020,Nichol2021}, also known as diffusion models.

VAEs are a stochastic variation of autoencoders (AEs), consisting of both an encoder and decoder. In VC methods employing VAEs \cite{Hsu2016short,Hsu2017short,vandenOord2017bshort,Huang2018,YSaito2018bshort,Kameoka2019IEEETransshort_ACVAE-VC,Tobing2019short,Tanaka2023}, the encoder's primary function is to convert the acoustic features of input speech into latent variables, while the decoder undertakes the reverse process. The fundamental concept behind these methods involves conditioning the decoder on a target speaker code, in addition to the latent variables, enabling it to learn to generate acoustic features that closely resemble those produced by the intended speaker while maintaining linguistic consistency with the input speech. Consequently, these methods have the capability to simultaneously learn mappings to multiple speakers' voices using a single pair of encoder and decoder networks.
Subsequent to these methods, regular AE-based methods, including AutoVC \cite{Qian2019}, AdaIN-VC \cite{Chou2019}, and AGAIN-VC \cite{Chen2021}, were introduced and demonstrated their ability to manage zero-shot any-to-any conversions. This was achieved by having the decoder utilize speaker embeddings generated by a speaker encoder, which was pretrained with the generalized end-to-end (GE2E) loss for speaker verification \cite{Wan2018}.

A non-parallel VC method was proposed using CycleGAN \cite{Kaneko2018,Kaneko2019short_cycleganvc2,Kaneko2020,Kaneko2021}, a variant of GAN originally designed for unpaired image translation \cite{Zhu2017short,Kim2017short,Yi2017short}. This approach involves training mappings between voices of different speakers using both cycle-consistency and adversarial loss. The adversarial loss ensures that the output follows the target distribution, while the cycle-consistency loss ensures that converting speech to another voice and back results in the original input. Remarkably, the effectiveness of the cycle-consistency loss extends to VAE-based methods as well \cite{Tobing2019short}. Despite the satisfactory performance of CycleGAN-based VC, its capability is confined to one-to-one conversions.
To address this constraint, an enhanced version \cite{Kameoka2018SLTshort_StarGAN-VC,Kaneko2019short_starganvc2,Kameoka2020IEEE-ASLP_StarGAN} based on another GAN variant known as StarGAN \cite{Choi2017short} was introduced subsequently. This approach combines the benefits of VAE-based and CycleGAN-based methods simultaneously. Similar to VAE-based techniques, it has the ability to learn mappings to multiple speakers' voices concurrently using a single network, allowing for the full utilization of available training data collected from various speakers. Moreover, it operates without requiring source speaker information, enabling it to handle any-to-many conversions effectively.

Flow-based models are generative models employing invertible layers known as flows \cite{Dinh2015,Dinh2017,Kingma2018}. These flows gradually transform real data samples into noise samples adhering to a specified distribution. Their aim is to directly evaluate and optimize the log-likelihood function using a special network architecture consisting of flows with easily computable Jacobians and inverse functions. A non-parallel VC method leveraging flow-based models was introduced \cite{Serra2019neurips}, resembling VAE-based methods where forward and inverse flows act akin to encoder and decoder functions.

Score-based generative models \cite{Song2019,Song2020}, also known as diffusion probabilistic models (DPMs) \cite{Ho2020,Nichol2021}, represent another noteworthy class of generative models distinct from those mentioned earlier. They have recently demonstrated remarkable effectiveness in generating both images and speech waveforms. Previously, we introduced VoiceGrad \cite{Kameoka2020arXiv_VoiceGrad,Kameoka2024_VoiceGrad}, a VC technique that enables mel-spectrogram conversion from a source to a target speaker using a score-based diffusion model. The concept entails training a score network to predict the gradient of the log density of mel-spectrograms from various speakers. Utilizing the trained score network, VC can be executed by employing annealed Langevin dynamics or reverse diffusion processes to iteratively adjust an input mel-spectrogram until it resembles one likely to be produced by the target speaker. 
An essential rationale for embracing a score-based generative model or DPM in VC is the ability to tailor the conversion process to suit diverse user needs. We envision achieving this by integrating independently pre-trained classifiers or other types of score-based models to modify the update direction during each time step of the Langevin dynamics or reverse diffusion process. This feature is especially attractive as it permits customization of converted speech without the need for additional training.
 Subsequently, several DPM-based VC methods, such as Diff-SVC \cite{Liu2021b}, Diff-VC \cite{Popov2022}, Diff-HierVC \cite{Choi2023}, and StableVC \cite{Yao2025} were proposed following the publication of the preprint paper on VoiceGrad. Diff-VC involves initially converting an input mel-spectrogram to an ``average voice'' mel-spectrogram using an encoder trained with phoneme supervision on speech samples from multiple speakers, and then to the target speaker's mel-spectrogram using the reverse diffusion process of a trained DPM. VoiceGrad distinguishes itself by directly initiating Langevin dynamics or reverse diffusion processes from the mel-spectrogram of the source speech. 

While VoiceGrad showed promising results, it faces several ongoing challenges. Firstly, there is significant scope for enhancing the audio quality of the converted speech. Secondly, the iterative nature of mel-spectrogram conversion tends to be slower compared to modern VC methods optimized for high-speed operation. One possible explanation for VoiceGrad's limited performance is the high dimensionality of mel-spectra (albeit with a much compressed representation compared to waveforms), making prediction of their score functions challenging and the conversion process time-consuming. To tackle these issues, this paper introduces the concept of latent diffusion models \cite{Rombach2022} to the VoiceGrad framework, proposing an enhanced version that integrates reverse diffusion in the low-dimensional autoencoder bottleneck.
Additionally, we explore using a flow matching model \cite{Lipman2023} as an alternative to the DPM to determine if it can further speed up the conversion process without compromising the quality of the converted speech.

\section{VoiceGrad}

\subsection{Diffusion probabilistic models}
\label{subsec:dpm}
We start by reviewing the concept of diffusion probabilistic models (DPMs) \cite{Ho2020}, or diffusion models for short.
DPMs are a type of generative model that produce data by reversing a gradual diffusion process. The diffusion process is usually defined by a sequence of noise addition steps, where each step slightly corrupts the data more than the previous one. Based on this definition, the reverse process is often modeled with a neural network that predicts the noise added at each step.

Formally, 
given a data sample $\Vec{x}_0$, normalized to have mean zero and unit variance,
the diffusion process in DPMs is defined as a Markov chain 
$q(\Vec{x}_{1:L}|\Vec{x}_0) = \prod_{l=1}^{L} q(\Vec{x}_{l}|\Vec{x}_{l-1})$ 
that produces a sequence $\Vec{x}_{1:L} = \{\Vec{x}_{1},\ldots,\Vec{x}_{L}\}$ from $\Vec{x}_0$
where 
$l=1,\ldots,L$ are the diffusion timesteps and 
$q(\Vec{x}_{l}|\Vec{x}_{l-1})$ is assumed to be a process of 
adding zero-mean Gaussian noise with variance $\beta_l$ to $\Vec{x}_{l-1}$
after scaling $\Vec{x}_{l-1}$ by $\sqrt{1-\beta_l}$
\begin{align}
q(\Vec{x}_{l}|\Vec{x}_{l-1})&=
\mathcal{N}(\Vec{x}_{l}; \sqrt{1-\beta_l}\Vec{x}_{l-1}, \beta_l\Vec{I}).
\end{align}
The reverse process, on the other hand, is modeled as a Markov chain 
$p_{\theta}(\Vec{x}_{0:L}) = p_{\theta}(\Vec{x}_L) \prod_{l=1}^{L} p_{\theta}(\Vec{x}_{l-1}|\Vec{x}_{l})$
starting from
$\Vec{x}_L\sim\mathcal{N}(\Vec{0},\Vec{I})$, where
$p_{\theta}(\Vec{x}_{l-1}|\Vec{x}_{l})$ is defined as
a Gaussian distribution with mean $\Vec{\mu}_{\theta}(\Vec{x}_l, l)$ 
and variance $\nu_l^2 \Vec{I}$
\begin{align}
p_{\theta}(\Vec{x}_{l-1}|\Vec{x}_{l})=
\mathcal{N}(
\Vec{x}_{l-1}; 
\Vec{\mu}_{\theta}(\Vec{x}_l, l),
\nu_l^2 \Vec{I}
).
\label{eq:DPM_5}
\end{align}
This corresponds to assuming the reverse diffusion step
as
\begin{align}
\Vec{x}_{l-1} =  \Vec{\mu}_{\theta}(\Vec{x}_l, l) + \nu_l \Vec{\epsilon},
\label{eq:reverse_step}
\end{align}
where $\Vec{\epsilon}$ is a random variable that follows a standard Gaussian.
By using $\alpha_l = 1-\beta_l$ and $\bar{\alpha}_l = \prod_{i=1}^{l} \alpha_{i}$, 
the mean
$\Vec{\mu}_{\theta}(\Vec{x}_l, l)$ is usually
reparametrized as 
\begin{align}
\Vec{\mu}_{\theta}(\Vec{x}_l, l)
=
\frac{1}{\sqrt{\alpha_l}}
\left(
\Vec{x}_l - 
\frac{
1 - \alpha_l
}{
\sqrt{1-\bar{\alpha}_l}
}
\Vec{\epsilon}_{\theta}(\Vec{x}_l, l)
\right),
\label{eq:DPM_7}
\end{align}
where $\Vec{\epsilon}_{\theta}(\Vec{x}_l, l)$ is modeled as a neural network with parameter $\theta$ conditioned on $l$.
This neural network is often called the score network because it corresponds to the score function of $\Vec{x}_0$.
The goal of training DPMs is to find $\theta$ such that 
the negative log-likelihood $\mathbb{E}[-\log p_{\theta}(\Vec{x}_0)]$ becomes as small as possible.
A typical way to achieve this is to search for $\theta$ by using a variational bound on $\mathbb{E}[-\log p_{\theta}(\Vec{x}_0)]$
\begin{multline}
\mathcal{L}_{\rm DPM}(\theta) = 
\mathbb{E}_{\Vec{x}_{1:L}\sim q(\Vec{x}_{1:L}|\Vec{x}_0)}
\left[
\log \frac{
q(\Vec{x}_{1:L}|\Vec{x}_0)
}{
p_{\theta}(\Vec{x}_{0:L})
}
\right]\\
=
\sum_{l=1}^{L}
\mathbb{E}_{\Vec{x}_0, \Vec{\epsilon}}
[
c_l
\|
\Vec{\epsilon}_{\theta}(
\sqrt{\bar{\alpha}_l} \Vec{x}_0 + \sqrt{1-\bar{\alpha}_l} \Vec{\epsilon},
l
)
-
\Vec{\epsilon}
\|_2^2
]
,
\label{eq:DPM_8}
\end{multline}
as the training objective to minimize,
where the expectation is taken over the training examples of $\Vec{x}_0$ and the random samples of $\Vec{\epsilon}\sim \mathcal{N}(\Vec{0},\Vec{I})$, and $c_l$ is a constant related to $\alpha_l$, $\bar{\alpha}_l$, and $\nu_l$.
Once $\theta$ is trained, a sample from $p_{\theta}(\Vec{x})$ can be generated using the Markov chain of the reverse process (recursion of \refeq{reverse_step}).

\subsection{Application to Voice Conversion}

The original VoiceGrad \cite{Kameoka2024_VoiceGrad} is based on the idea of performing a reverse diffusion process starting from a source mel-spectrogram, assuming it is a diffused version of a target mel-spectrogram at a certain timestep $L'$. The score network is designed to be conditioned on the speaker embedding $\Vec{s}$, timestep $l$, and phoneme embedding sequence $\Vec{p}$, and is trained using speech samples from various speakers. This allows us to transform the mel-spectrogram of the source speech to resemble that of a target speaker by using phoneme embeddings extracted from the source speech and a speaker embedding extracted from the target speaker as cues.
The original VoiceGrad used a one-hot vector for the speaker embedding, limiting it to converting input speech to the voices of known speakers. In this paper, we modify it to use the speaker embedding generated by a pre-trained GE2E speaker encoder \cite{Wan2018} from a reference voice, enabling zero-shot conversion (i.e., any-to-any conversion). 
For the phoneme embedding sequence $\Vec{p}$, VoiceGrad utilized the feature sequence extracted by the phoneme feature extractor proposed by Liu et al. \cite{Liu2021}.  
This feature extractor is created by inserting an additional layer between the encoder and decoder in an end-to-end phoneme recognizer \cite{Kim2017}, training the entire network on a large speech recognition dataset, and then removing all layers following the inserted layer after training.

In the original VoiceGrad, using the $L_1$ measure instead of the $L_2$ measure in \refeq{DPM_8} proved effective for both training stability and audio quality at test time. 
Therefore, ignoring the constant terms, the training objective becomes
\begin{align}
\mathcal{L}_{\rm VG1}(\theta) = 
\mathbb{E}
[
\|
\Vec{\epsilon}_{\theta}(
\sqrt{\bar{\alpha}_l} \Vec{x}_0 + \sqrt{1-\bar{\alpha}_l} \Vec{\epsilon},
l,
\Vec{s},
\Vec{p}
)
-
\Vec{\epsilon}
\|_1
],
\label{eq:L_VG1}
\end{align}
where 
$\Vec{x}_0$ represents the mel-spectrogram of a training utterance, 
the expectation is taken over
$l\sim \mathcal{U}[1,\ldots,L]$,
$k \sim \mathcal{U}[1,\ldots,K]$, 
$\Vec{\epsilon} \sim \mathcal{N}(\Vec{0},\Vec{I})$,
and
$\Vec{x}_0 \sim p(\Vec{x}|k)$,
$k$ denotes a speaker index,
$K$ denotes the number of speakers included in the training set,
$\mathcal{U}[\cdot]$ represents a discrete uniform distribution over its specified range,
$p(\Vec{x}|k)$ represents the distribution of the mel-spectrograms of speaker $k$,
and 
$\Vec{s}$ and $\Vec{p}$ represent the speaker embedding and phoneme embedding sequence, respectively,
extracted from $\Vec{x}_0$.

Let $\mathsf{SpeakerEncoder}(\cdot)$ and $\mathsf{PhonemeEncoder}(\cdot)$ represent the same speaker and phoneme encoders used to extract $\Vec{s}$ and $\Vec{p}$ during training.
Let $\Vec{x}$ and $\Vec{x}_{\rm ref}$ denote the mel-spectrograms of the source speech to be converted and the reference speech of a target speaker, respectively.
The VC process can then be described as shown in \refalgo{VoiceGrad}, where $L'$ denotes the starting noise level of the iteration and $\Vec{h}$ represents a vocoder that finally transforms the converted $\Vec{x}$ into a waveform $\Vec{y}$. 
\begin{algorithm}[t!]
\caption{VoiceGrad (DPM version) VC process}
\label{algo:VoiceGrad}
\begin{algorithmic}
\REQUIRE $\{\alpha_l\}_{l=1}^{L}$, $\{\bar{\alpha}_l\}_{l=1}^{L}$, $\Vec{x}$, $\Vec{x}_{\rm ref}$, $L'$
\STATE $\Vec{s}\leftarrow \mathsf{SpeakerEncoder}(\Vec{x}_{\rm ref})$
\STATE $\Vec{p}\leftarrow \mathsf{PhonemeEncoder}(\Vec{x})$
\FOR{$l=L'$ to $1$}
\STATE Draw $\Vec{\epsilon} \sim\mathcal{N}(\Vec{0},\Vec{I})$
\STATE Update 
$\Vec{x}
\leftarrow
\frac{1}{\sqrt{\alpha_l}}
\left(
\Vec{x} - 
\frac{
1 - \alpha_l
}{
\sqrt{1-\bar{\alpha}_l}
}
\Vec{\epsilon}_{\theta}(\Vec{x}, l, \Vec{s}, \Vec{p})
\right) + \nu_l \Vec{\epsilon}$
\ENDFOR
\STATE $\Vec{y}\leftarrow \Vec{h}(\Vec{x})$
\RETURN $\Vec{y}$
\end{algorithmic}
\end{algorithm}

\section{Proposed: LatentVoiceGrad}

\subsection{Basic Idea}

\begin{figure*}[t!]
\centering
\begin{minipage}[t!]{.85\linewidth}
  \centerline{\includegraphics[width=.98\linewidth]{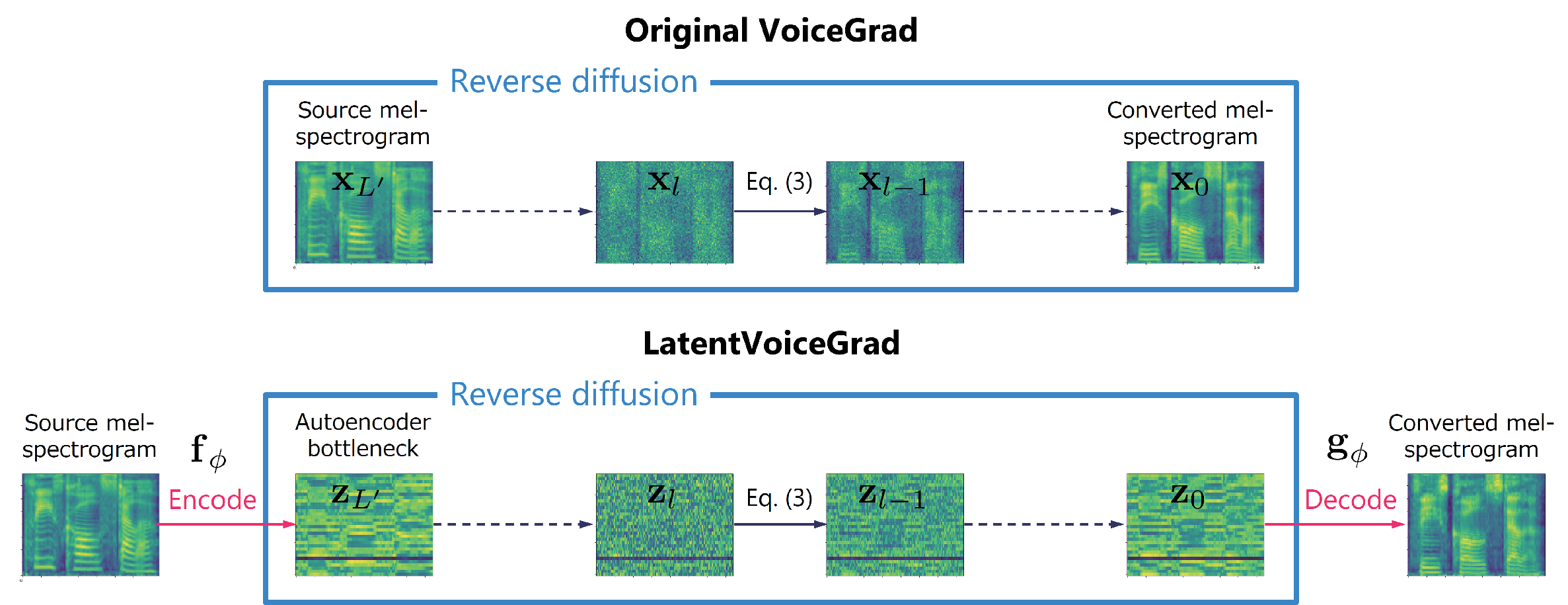}}
  \vspace{-0ex}
  \centering
  \caption{Concepts of the DPM versions of VoiceGrad and LatentVoiceGrad.}
\label{fig:latentvoicegrad}
\end{minipage}
\end{figure*}

Although the original VoiceGrad performed reasonably well, it faced a tradeoff between conversion quality and inference speed, requiring about 10 to 20 iterations to achieve satisfactory conversion.
To speed up the conversion process, we propose applying a reverse diffusion process to a lower-dimensional latent representation of mel-spectrograms, inspired by the success of the latent diffusion models \cite{Rombach2022} in image generation tasks.
If a good lower-dimensional latent space can be found, we would expect not only faster inference but also that predicting the score function in that space will be relatively less challenging due to its lower dimensionality.
To obtain a lower-dimensional latent representation of mel-spectrograms, we train a speaker-independent undercomplete autoencoder with a bottleneck structure, consisting of an encoder and decoder, on speech samples from various speakers.
As will be described later, the autoencoder is trained using an adversarial loss, with the goal of encouraging its output to ultimately produce high-quality waveforms when passed through the vocoder it is designed to work with.
After training the autoencoder, its encoder outputs (i.e., the autoencoder bottleneck features) are treated as the data to be modeled by the DPM.
The new version of VoiceGrad that incorporates this idea will be called {\it LatentVoiceGrad} (\reffig{latentvoicegrad}).
In this paper, we also explore using a flow matching model \cite{Lipman2023} as an alternative to the DPM to potentially further speed up conversion without compromising audio quality.

While conventional autoencoder-based VC methods often integrate data transformation capabilities into the autoencoder itself, our method deliberately avoids this. Instead, we use an undercomplete autoencoder solely for compressing and reconstructing mel-spectrograms. This design choice reflects our belief that undercomplete autoencoders are well-suited for compression and reconstruction but are less effective than diffusion and flow models when it comes to data transformation. Conversely, diffusion and flow models are highly effective at transforming low-dimensional data but tend to struggle with high-dimensional data, both in terms of conversion speed and output quality. LatentVoiceGrad is designed to leverage the strengths of both components while compensating for their individual limitations (\reftab{roles_and_trade-offs}).

\begin{table*}[t!]
\centering
\caption{Roles and trade-offs of autoencoder and flow/diffusion models in LatentVoiceGrad.}
\label{tab:roles_and_trade-offs}
\begin{tabular}{@{} p{2.5cm} p{2.5cm} p{3.5cm} p{3.8cm} p{4cm} @{}}
\toprule
\textbf{Component} & \textbf{Conventional Use} & \textbf{Strengths} & \textbf{Weaknesses} & \textbf{Role in Proposed Method} \\ 
\midrule
Autoencoder & Compression \& optional data transformation & Excels at compressing and reconstructing mel-spectrograms & Limited capability for data transformation & Used exclusively for compression and reconstruction \\ 
Flow/Diffusion Model & Data transformation & Well-suited for transforming low-dimensional data & Struggles with high-dimensional data (slow conversion, lower quality) & Used for transforming compressed latent representations \\ 
LatentVoiceGrad (Proposed Method) & Combines both & Leverages strengths of autoencoder and flow/diffusion model & --- & Achieves effective conversion by combining components while mitigating their weaknesses \\ 
\bottomrule
\end{tabular}
\end{table*}

\subsection{Adversarial Autoencoder Training}
\label{subsec:aegan}

Since we want the autoencoder to generate mel-spectrograms that will eventually be transformed into realistic audio waveforms, we train it using both a regular reconstruction loss and an adversarial loss that evaluates the signals generated by a pretrained, fixed vocoder. Specifically, we train the autoencoder alongside a waveform discriminator, so the autoencoder improves at generating mel-spectrograms from which a pretrained vocoder can produce realistic waveforms, while the discriminator improves at distinguishing those generated waveforms as fake data.
Additionally, since the DPM assumes the data samples to be zero-centered and have unit variance, we want the autoencoder bottleneck to satisfy this condition as well. To achieve this, we include a Kullback-Leibler (KL) divergence between the distribution of the bottleneck features and a standard normal distribution in the training loss, similar to VAE training.

In the following, we use $\Vec{y}$ and $\Vec{x}$ to denote the waveform and its mel-spectrogram of a training speech sample, respectively. 
Using $\Vec{g}_{\phi}(\cdot)$ and $\Vec{f}_{\phi}(\cdot)$ to denote the encoder and decoder of the autoencoder with parameter $\phi$, the regular reconstruction loss $\mathcal{L}_{\rm rec}(\phi)$ is defined as an L1 loss in the mel-spectrogram domain:
\begin{align}
\mathcal{J}_{\rm rec}(\phi) = 
\mathbb{E}_{\Vec{x}}
[
\|
\Vec{g}_{\phi}(\Vec{f}_{\phi}(\Vec{x}))
- \Vec{x}
\|_1
].
\end{align}
For the waveform discriminator, we use a combination of the multi-period discriminator (MPD) and multi-scale discriminator (MSD) introduced in HiFi-GAN \cite{Kong2020} and MelGAN \cite{Kumar2019}. 
Using $d_{\psi}(\cdot)\in [0,1]$ to denote the MPD and MSD treated as a single discriminator with parameter $\psi$, the adversarial losses for the autoencoder and discriminator, $\mathcal{L}_{\rm adv}^{\rm (a)}(\phi)$ and $\mathcal{L}_{\rm adv}^{\rm (d)}(\psi)$, are defined as the least squares GAN objectives \cite{Mao2017}
\begin{align}
\mathcal{J}_{\rm adv}^{\rm (a)}(\phi) &=
\mathbb{E}_{\Vec{x}}
[
(
d_{\psi}(
\hat{\Vec{y}}
)
-1
)^2
],
\label{eq:adversarial_loss_gen}
\\
\mathcal{J}_{\rm adv}^{\rm (d)}(\psi) &=
\mathbb{E}_{\Vec{y}}
[
(
d_{\psi}(\Vec{y})
-1
)^2
]
+
\mathbb{E}_{\Vec{x}}
[
(
d_{\psi}(
\hat{\Vec{y}}
)
)^2
],
\label{eq:adversarial_loss_dis}
\end{align}
where 
$\Vec{h}(\cdot)$ represents a pretrained (frozen) neural vocoder that transforms a mel-spectrogram into a waveform and
$\hat{\Vec{y}} = \Vec{h}(\Vec{g}_{\phi}(\Vec{f}_{\phi}(\Vec{x}))$ is a waveform generated by the neural vocoder from the autoencoder output.
In this paper, we used the pretrained universal HiFi-GAN vocoder taken from the official HiFi-GAN repository\footnote{https://github.com/jik876/hifi-gan}
for $\Vec{h}(\cdot)$.
Minimizing \refeq{adversarial_loss_dis} with respect to $\psi$
means training the discriminator to classify real speech samples as 1 and the samples synthesized by the generator (i.e., autoencoder and vocoder) as 0. 
Conversely, minimizing \refeq{adversarial_loss_gen} with respect to $\phi$ means training the autoencoder to deceive the discriminator, i.e., to make the discriminator classify the synthesized samples as 1.
As in the HiFi-GAN training, we further incorporate the mel-spectrogram and feature matching losses 
\begin{align}
\mathcal{J}_{\rm mel}(\phi) &=  
\mathbb{E}_{\Vec{x}}
[
\|
\Vec{m}(
\hat{\Vec{y}}
) 
- 
\Vec{x}
\|_1
],
\\
\mathcal{J}_{\rm feat}(\phi) &= 
\mathbb{E}_{(\Vec{x},\Vec{y})}
\left[
\sum_{i=1}^{T}
\frac{1}{N_i}
\|
d_{\psi}^{(i)}(
\hat{\Vec{y}}
)
-
d_{\psi}^{(i)}(\Vec{y})
\|_1
\right],
\end{align}
in the training objective for the autoencoder, where 
$\Vec{m}(\cdot)$ represents the process of computing the mel-spectrogram from a waveform,
$T$ is the number of the layers in the discriminator,
and $d_{\psi}^{(i)}(\cdot)$ and $N_i$ denote the features and the number of features in the $i$-th layer of the discriminator, respectively.

The KL loss $\mathcal{J}_{\rm KL}(\phi)$ to regularize the latent $\Vec{z} = \Vec{g}_{\phi}(\Vec{x})$ to be zero-centered and have unit variance is defined as 
\begin{align}
\mathcal{J}_{\rm KL}(\phi) = 
\frac{1}{2}
\mathbb{E}_{\Vec{x}}
[
\sigma^2(\Vec{x}) + \mu(\Vec{x})^2 - 1 - \log \sigma^2(\Vec{x})
],
\end{align}
where $\mu(\Vec{x})$ and $\sigma^2(\Vec{x})$ denote the mean and variance of the elements of $\Vec{z}= \Vec{g}_{\phi}(\Vec{x})$, respectively.

Putting it all together, the training objectives to minimize with respect to $\phi$ and $\psi$ are given as
$\mathcal{J}_{\rm ae}(\phi) = 
\lambda (
\mathcal{J}_{\rm rec}(\phi) + 
\mathcal{J}_{\rm mel}(\phi) +
\mathcal{J}_{\rm KL}(\phi)
) +
\mathcal{J}_{\rm adv}^{\rm (a)}(\phi) +
\mathcal{J}_{\rm feat}(\phi)$
and
$\mathcal{J}_{\rm dis}(\psi) = \mathcal{J}_{\rm adv}^{\rm (d)}(\psi)$,
respectively, where $\lambda$ is a regularization parameter that weighs the importance of $\mathcal{J}_{\rm rec}(\phi)$, $\mathcal{J}_{\rm mel}(\phi)$, and $\mathcal{J}_{\rm KL}(\phi)$.
In this paper, we set $\lambda$ to 45.

In the current experiment, the channel numbers of the autoencoder bottleneck and the mel-spectrogram were set to 32 and 80, respectively.
The architectures of the encoder and decoder were each implemented as a stack consisting of a fully connected layer (with 256 output channels) followed by a leaky rectified linear unit (LReLU) activation, two-layer bidirectional long short-term memory (BiLSTM) networks (with 256 output channels), and another fully connected layer (with 32 output channels for the encoder and 80 output channels for the decoder).

\subsection{DPM Training}

While VoiceGrad works by gradually transforming a source mel-spectrogram into a target mel-spectrogram through a reverse diffusion process, LatentVoiceGrad uses the autoencoder bottleneck features of the source mel-spectrogram as the data to be transformed (\reffig{latentvoicegrad}). 
Once the autoencoder has been trained as described above, the score network is trained on the autoencoder bottleneck features extracted from the training speech samples of various speakers. Like in VoiceGrad, the score network is conditioned on the speaker embedding $\Vec{s}$, timestep $l$, and phoneme embedding sequence $\Vec{p}$.
If we use $\Vec{z}_0$ to denote the autoencoder bottleneck features extracted from the mel-spectrogram $\Vec{x}_0$ of a training utterance, the training objective remains the same as in \refeq{L_VG1}, but with $\Vec{x}_0$ replaced by $\Vec{z}_0$.
After the score network is trained, input speech can be converted to a target speaker's voice by transforming its audio signal into the bottleneck features using the encoder $\Vec{f}_{\phi}$, modifying them through the reverse diffusion process using the target speaker embedding $\Vec{s}$, and finally generating a waveform using the decoder $\Vec{g}_{\phi}$ and vocoder $\Vec{h}$ (\refalgo{LatentVoiceGrad}).
\begin{algorithm}[t!]
\caption{LatentVoiceGrad (DPM version) VC process}
\label{algo:LatentVoiceGrad}
\begin{algorithmic}
\REQUIRE $\{\alpha_l\}_{l=1}^{L}$, $\{\bar{\alpha}_l\}_{l=1}^{L}$, $\Vec{x}$, $\Vec{x}_{\rm ref}$, $L'$
\STATE $\Vec{s}\leftarrow \mathsf{SpeakerEncoder}(\Vec{x}_{\rm ref})$
\STATE $\Vec{p}\leftarrow \mathsf{PhonemeEncoder}(\Vec{x})$
\STATE $\Vec{z}\leftarrow \Vec{f}_{\phi}(\Vec{x})$
\FOR{$l=L'$ to $1$}
\STATE Draw $\Vec{\epsilon} \sim\mathcal{N}(\Vec{0},\Vec{I})$
\STATE Update 
$\Vec{z}
\leftarrow
\frac{1}{\sqrt{\alpha_l}}
\left(
\Vec{z} - 
\frac{
1 - \alpha_l
}{
\sqrt{1-\bar{\alpha}_l}
}
\Vec{\epsilon}_{\theta}(\Vec{z}, l, \Vec{s}, \Vec{p})
\right) + \nu_l \Vec{\epsilon}$
\ENDFOR
\STATE $\Vec{y}\leftarrow \Vec{h}(\Vec{g}_{\phi}(\Vec{z}))$
\RETURN $\Vec{y}$
\end{algorithmic}
\end{algorithm}

\subsection{Flow Matching Approach}

When the number of timesteps in DPMs approaches infinity, their diffusion process can be understood as a stochastic differential equation (SDE). In this view, sampling data via reverse diffusion corresponds to numerically solving the reverse-time SDE or the associated probability-flow ordinary differential equation (ODE).
Recently, a new class of generative models known as flow matching (FM) models \cite{Lipman2023} has emerged as an advanced version of continuous normalizing flows (CNF) \cite{Chen2018}, 
and there has been a growing interest in exploring their application to VC tasks \cite{Chen2024_SLT, Yao2025}. 
Unlike DPMs, which learn the score function of an SDE, FM models aim to directly learn the vector field of an arbitrary ODE. They use a neural network to approximate this vector field, which can then be used to numerically solve the ODE to generate data samples. These models often design their ODEs and vector fields to form straight line sampling trajectories and minimize transport costs, resulting in simpler, more efficient formulations and better-quality outcomes. 
Given this advantage, we investigate using an FM model as an alternative to the DPM to see if it can further accelerate the VoiceGrad conversion process without compromising speech quality.

Here, we review the origins and rationale behind the FM concept.
Let $\Vec{x}_1$ represent a data sample generated according to an unknown data distribution $q_1(\Vec{x})$. 
FM models, like CNF models, assume that $\Vec{x}_1$ is obtained by gradually transforming an initial point $\Vec{x}_0$ in the data space through an ODE:
\begin{align}
\frac{\mbox{d}\Vec{x}_t}{\mbox{d}t}  = 
\Vec{u}_t(\Vec{x}_t),
\label{eq:ode}
\end{align}
where $t$ represents time starting from 0 and ending at 1, 
$\Vec{x}_t$ represents the value at time $t$,
and
$\Vec{u}_t$ represents the vector field governing the dynamics of the ODE.
Note that in the description of DPM in \refsubsec{dpm}, the data sample was denoted as $\Vec{x}_0$, whereas here it is denoted as $\Vec{x}_1$, following the literature on FM models, with the initial point of the ODE denoted as $\Vec{x}_0$.
The core concept in CNF and FM models, which builds on neural ODEs \cite{Chen2018}, is to represent $\Vec{u}_t(\Vec{x})$ using a neural network $\Vec{v}_{\theta}(\Vec{x}, t)$ with parameter $\theta$.
If the uniqueness of the solution of \refeq{ode} is satisfied,
\refeq{ode} defines a diffeomorphism
$\Vec{x}_t = \Vec{\phi}_t(\Vec{x}_0) = \Vec{x}_0 + \int_{0}^{t} \Vec{u}_s(\Vec{x}_s) \mbox{d}s$, called a flow. 
This means that once $\theta$ is obtained, data sampling can be performed by solving \refeq{ode} with $\Vec{u}_t(\Vec{x})$ replaced by $\Vec{v}_{\theta}(\Vec{x}, t)$, specifically by  numerically computing $\Vec{x}_1 = \Vec{\phi}_1(\Vec{x}_0)$, using methods such as Euler's method.
Therefore, our focus is on how to train $\theta$ using the training examples of $\Vec{x}_1\sim q_1(\Vec{x})$.

Assuming $\Vec{x}_0$ is a sample drawn from a simple distribution $q_0$ (e.g., a standard normal distribution), the density $p_t$ of $\Vec{x}_t$ at time $t$, called a probability path, can be described using $q_0$ and $\Vec{\phi}_t$ via the change of variables formula (or push-forward equation). 
To determine if a given vector field $\Vec{v}_{\theta}$ can generate a valid probability path $p_t$, one practical approach is to test if $\Vec{v}_{\theta}$ and $p_t$ satisfy the continuity equation
\begin{align}
\frac{\mbox{d}p_t(\Vec{x})}{\mbox{d}t} + \nabla \cdot (p_t(\Vec{x}) \Vec{v}_{\theta}(\Vec{x},t)) = 0.
\label{eq:continuity_equation}
\end{align}
Under this constraint, the goal is to find $\theta$ such that $p_1=q_1$. 
In CNF models, $\theta$ is trained to directly maximize the log-likelihood $\mathbb{E}_{\Vec{x}_1 \sim q_1(\Vec{x})} [\log p_1(\Vec{x}_1)]$, which can be derived from \refeq{continuity_equation}. However, this approach can be computationally very expensive because both evaluating and differentiating $\log p_1(\Vec{x}_1)$ require integration over time for each individual sample.
Instead of using the log-likelihood, if we know the real vector field $\Vec{u}_t(\Vec{x})$, we might want to use a loss function that directly measures the distance between $\Vec{v}_{\theta}(\Vec{x},t)$ and $\Vec{u}_t(\Vec{x})$:
\begin{align}
\mathcal{L}_{\rm FM}(\theta) = \mathbb{E}_{t, \Vec{x}\sim p_t(\Vec{x})}[
\|
\Vec{v}_{\theta}(\Vec{x},t) - \Vec{u}_t(\Vec{x})
\|_2^2
],
\label{eq:L_FM}
\end{align}
as the training objective instead.
This loss is appealing because it avoids integration over $t$ and can be evaluated at each $t$ independently.
Unfortunately, this loss is intractable to use in practice because both $\Vec{u}_t$ and $p_t$ are not directly accesible.
However, an interesting fact is that if we express $p_t(\Vec{x})$ as a mixture of ``conditional probability paths'' $p_t(\Vec{x}|\Vec{c})$ that vary according to some conditioning variable $\Vec{c}$
\begin{align}
p_t(\Vec{x}) = \int p_t(\Vec{x}|\Vec{c}) \pi(\Vec{c}) \mbox{d}\Vec{c},
\label{eq:mixture_of_conditional_probability_paths}
\end{align}
a loss function defined as
\begin{multline}
\mathcal{L}_{\rm CFM}(\theta) 
\\
=
\mathbb{E}_{t, \Vec{c}\sim \pi(\Vec{c}), \Vec{x}\sim p_t(\Vec{x}|\Vec{c})}[
\|
\Vec{v}_{\theta}(\Vec{x},t) - \Vec{u}_t(\Vec{x}|\Vec{c})
\|_2^2
],
\label{eq:L_CFM}
\end{multline}
where 
$\pi$ represents the distribution over $\Vec{c}$ and 
$\Vec{u}_t(\Vec{x}|\Vec{c})$ represents the vector field associated with $p_t(\Vec{x}|\Vec{c})$,
is equal to $\mathcal{L}_{\rm FM}(\theta)$ up to a constant independent of $\theta$.
While this holds for arbitrary $\Vec{c}$, 
one convenient assumption for $p_t(\Vec{x}|\Vec{c})$ is given by Tong et al. \cite{Tong2024}, who propose putting $\Vec{c}=(\Vec{x}_0, \Vec{x}_1)$ and assuming $p_t(\Vec{x}|\Vec{x}_0, \Vec{x}_1)$ and $\pi(\Vec{x}_0, \Vec{x}_1)$ as
\begin{align}
p_t(\Vec{x}|\Vec{x}_0, \Vec{x}_1) &= 
\mathcal{N}(\Vec{x} | t \Vec{x}_1 + (1-t)\Vec{x}_0, \sigma^2 \Vec{I}),
\label{eq:Gaussian_conditional_probability_path}
\\
\pi(\Vec{x}_0, \Vec{x}_1) &=
q_0(\Vec{x}_0)
q_1(\Vec{x}_1),
\label{eq:pi(z)}
\end{align}
where $\sigma^2$ is a manually set constant.
Under these settings, 
it can be confirmed from \refeq{mixture_of_conditional_probability_paths} that
$p_1(\Vec{x})$
becomes
\begin{align}
p_1(\Vec{x}) = 
\int 
q_1(\Vec{x}_1)
\mathcal{N}(\Vec{x} | \Vec{x}_1, \sigma^2 \Vec{I}) 
\mbox{d}\Vec{x}_1 \approx q_1(\Vec{x}),
\end{align}
which is a Gaussian mixture with each Gaussian centered at an individual training example $\Vec{x}_1$. 
\refeq{Gaussian_conditional_probability_path} 
implies that 
$p_t(\Vec{x}|\Vec{x}_0, \Vec{x}_1)$ is represented as a Gaussian with variance $\sigma^2$ whose center moves in a straight line from $\Vec{x}_0$ to $\Vec{x}_1$ at a constant velocity as $t$ varies from 0 to 1.
This corresponds to assuming $\Vec{x}_t$ as
\begin{align}
\Vec{x}_t =  t \Vec{x}_1 + (1-t)\Vec{x}_0 + \sigma \Vec{\epsilon},
\end{align}
where $\Vec{\epsilon}$ is a random variable that follows a standard Gaussian.
Hence, from \refeq{ode}, the vector field $\Vec{u}_t(\Vec{x}|\Vec{x}_0, \Vec{x}_1)$ associated with the assumed $p_t(\Vec{x}|\Vec{x}_0, \Vec{x}_1)$ is given explicitly as
\begin{align}
\Vec{u}_t(\Vec{x}|\Vec{x}_0, \Vec{x}_1) = 
\frac{\mbox{d}\Vec{x}_t}{\mbox{d}t} = 
\Vec{x}_1 - \Vec{x}_0.
\label{eq:conditional_vector_field}
\end{align}
Assuming $q_0=\mathcal{N}(\Vec{0},\Vec{I})$ and
plugging (\ref{eq:Gaussian_conditional_probability_path}), (\ref{eq:pi(z)}), and (\ref{eq:conditional_vector_field})
into \refeq{L_CFM}, $\mathcal{L}_{\rm CFM}$ can be written as 
\begin{align}
\mathcal{L}_{\rm CFM}(\theta) 
= 
\mathbb{E}
[
\|
\Vec{v}_{\theta}(\Vec{x}_t,t) - 
(\Vec{x}_1 - \Vec{x}_0)
\|_2^2
],
\label{eq:L_CFM2}
\end{align}
where the expectation is taken over 
$t \sim \mathcal{U}(0,1)$,
$\Vec{x}_0 \sim \mathcal{N}(\Vec{x}|\Vec{0},\Vec{I})$,
$\Vec{x}_1 \sim q_1(\Vec{x})$, and
$\Vec{x}_t \sim \mathcal{N}(\Vec{x}| t \Vec{x}_1 + (1-t)\Vec{x}_0, \sigma^2 \Vec{I})$,
with $\mathcal{U}(0,1)$ representing a continuous uniform distribution over the range $(0,1)$.
Notably, unlike the objective function in \refeq{L_FM}, this objective can be easily evaluated and differentiated with respect to $\theta$.

In DPM, the score network is trained to predict the noise added at each time step. The negative of this noise indicates the direction in which the data samples should be updated during the reverse diffusion process. 
DPM and FM are conceptually similar in this respect because the vector field $\Vec{v}_{\theta}$ represents the direction in which the data sample should be updated at each step of the ODE solving process.
The difference is that \refeq{L_CFM2} ensures the vector fields at all points in the neighborhood of the trajectory from the initial point $\Vec{x}_0$ to the real data sample $\Vec{x}_1$ are oriented in the direction of $\Vec{x}_1 - \Vec{x}_0$ during training. This makes the trajectories of the ODE straight, allowing the ODE solving process to be more efficient than the reverse diffusion process in DPM.  
Given the similarity between the two concepts and the advantage of FM, we can consider another version of VoiceGrad that replaces DPM with FM. Namely, just as the score network conditioned on a target speaker embedding in VoiceGrad can convert the mel-spectrogram of source speech to the target speaker's voice through the reverse diffusion process, it is expected that the vector field conditioned on the target speaker embedding can achieve the same more efficiently through the ODE solving process.

In the FM version of VoiceGrad, the vector field network $\Vec{v}_{\theta}$ is designed to be conditioned on the speaker embedding $\Vec{s}$, time $t$, and phoneme embedding sequence $\Vec{p}$. 
It is then trained using speech samples from various speakers using the $L_1$ measure instead of the $L_2$ measure
\begin{align}
\mathcal{L}_{\rm CFM}(\theta) 
= 
\mathbb{E}
[
\|
\Vec{v}_{\theta}(\Vec{x}_t, t, \Vec{s}, \Vec{p}) - 
(\Vec{x}_1 - \Vec{x}_0)
\|_1
],
\label{eq:L_VG2}
\end{align}
as the training objective, where the expectation is taken over 
$t\sim \mathcal{U}(0,1)$,
$k \sim \mathcal{U}[1,\ldots,K]$,
$\Vec{x}_0 \sim \mathcal{N}(\Vec{x}|\Vec{0},\Vec{I})$,
$\Vec{x}_1 \sim q_1(\Vec{x}|k)$, and
$\Vec{x}_t \sim \mathcal{N}(\Vec{x}|t \Vec{x}_1 + (1-t)\Vec{x}_0, \sigma^2 \Vec{I})$, 
$\Vec{x}_1$ represents the mel-spectrogram of a training utterance, 
$k$ represents a speaker index,
$q_1(\Vec{x}|k)$ represents the distribution of the mel-spectrograms of speaker $k$, 
$K$ is the number of speakers included in the training set,
$\Vec{s}$ represents the speaker embedding extracted from $\Vec{x}_1$, 
and $\Vec{p}$ represents the phoneme embedding sequence extracted from $\Vec{x}_1$.

The FM version can also follow a similar approach to LatentVoiceGrad by treating the autoencoder bottleneck as the data to be transformed, training the vector field network instead of the score network, and using an ODE solver instead of reverse diffusion to perform VC. 
In this case, if we use $\Vec{z}_1$ to denote the autoencoder bottleneck features extracted from the mel-spectrogram $\Vec{x}_1$ of a training utterance, the training objective will be the same as in \refeq{L_VG2}, but with $\Vec{x}_1$ replaced by $\Vec{z}_1$.
This results in four possible approaches, depending on whether we use the mel-spectrogram or the autoencoder bottleneck as the data to be converted, and whether the DPM or FM model serves as the underlying generative model.
We will refer to these as VoiceGrad-DPM, LatentVoiceGrad-DPM, VoiceGrad-FM, and LatentVoiceGrad-FM, and compare them experimentally in the following section. 
The VC processes of VoiceGrad-FM and LatentVoiceGrad-FM using Euler's method are outlined in \refalgos{VoiceGrad-FM}{LatentVoiceGrad-FM}, respectively.
Both algorithms were found to perform well when random noise was added to the initial point, using $\Vec{x}\leftarrow (1-r) \Vec{x} + r \Vec{\epsilon}$ or $\Vec{z}\leftarrow (1-r) \Vec{z} + r \Vec{\epsilon}$, where $\Vec{\epsilon} \sim\mathcal{N}(\Vec{0},\Vec{I})$ represents Gaussian noise and $0\le r\le 1$ represents the proportion of noise added.
\begin{algorithm}[t!]
\caption{VoiceGrad (FM version) VC process}
\label{algo:VoiceGrad-FM}
\begin{algorithmic}
\REQUIRE $\Vec{x}$, $\Vec{x}_{\rm ref}$, $r$, $L$
\STATE $\Vec{s}\leftarrow \mathsf{SpeakerEncoder}(\Vec{x}_{\rm ref})$
\STATE $\Vec{p}\leftarrow \mathsf{PhonemeEncoder}(\Vec{x})$
\STATE Draw $\Vec{\epsilon} \sim\mathcal{N}(\Vec{0},\Vec{I})$
\STATE $\Vec{x}\leftarrow (1-r) \Vec{x} + r \Vec{\epsilon}$
\FOR{$l=1$ to $L$}
\STATE Update 
$\Vec{x}
\leftarrow 
\Vec{x} + 
\frac{1}{L}
\Vec{v}_{\theta}(\Vec{x}, l/L, \Vec{s}, \Vec{p})
$
\ENDFOR
\STATE $\Vec{y}\leftarrow \Vec{h}(\Vec{x})$
\RETURN $\Vec{y}$
\end{algorithmic}
\end{algorithm}

\begin{algorithm}[t!]
\caption{LatentVoiceGrad (FM version) VC process}
\label{algo:LatentVoiceGrad-FM}
\begin{algorithmic}
\REQUIRE $\Vec{x}$, $\Vec{x}_{\rm ref}$, $r$, $L$
\STATE $\Vec{s}\leftarrow \mathsf{SpeakerEncoder}(\Vec{x}_{\rm ref})$
\STATE $\Vec{p}\leftarrow \mathsf{PhonemeEncoder}(\Vec{x})$
\STATE $\Vec{z}\leftarrow \Vec{f}_{\phi}(\Vec{x})$
\STATE Draw $\Vec{\epsilon} \sim\mathcal{N}(\Vec{0},\Vec{I})$
\STATE $\Vec{z}\leftarrow (1-r) \Vec{z} + r \Vec{\epsilon}$
\FOR{$l=1$ to $L$}
\STATE Update 
$\Vec{z}
\leftarrow 
\Vec{z} + 
\frac{1}{L}
\Vec{v}_{\theta}
(\Vec{z}, l/L, \Vec{s}, \Vec{p})
$
\ENDFOR
\STATE $\Vec{y}\leftarrow \Vec{h}(\Vec{g}_{\phi}(\Vec{z}))$
\RETURN $\Vec{y}$
\end{algorithmic}
\end{algorithm}

\section{Experiments}
\label{sec:experiments}

\subsection{Dataset}
\label{subsec:dataset}

To evaluate the performance of the four approaches---VoiceGrad-DPM, LatentVoiceGrad-DPM, VoiceGrad-FM, and LatentVoiceGrad-FM---we conducted zero-shot any-to-any VC experiments.
For the experiments, we used the CSTR VCTK Corpus (version 0.92)\footnote{\url{https://datashare.ed.ac.uk/handle/10283/3443}}, which contains speech data from 110 English speakers with various accents, each reading approximately 400 sentences.
To simulate a zero-shot 
{\it unseen-to-unseen}
VC scenario, i.e., an any-to-any VC scenario in which both the source and target speakers are excluded from the training data,
we used utterances from ten speakers (p238, p241, p243, p252, p261, p294, p334, p343, p360, and p362) as source and reference speech and used utterances from the remaining 100 speakers as training data.
This setup resulted in 90 source-target speaker combinations for testing, with the task being to convert input speech into a voice similar to that of reference speech. 
To ensure that the target speaker's reference speech and the source speech were different sentences, we used utterance number 003 of the test speaker's speech as the reference speech and the remaining utterances as the source speech to be converted.
All the speech signals were sampled at 22.05 kHz.

\subsection{Baseline Method}

We selected the original VoiceGrad (`VoiceGrad-DPM') \cite{Kameoka2024_VoiceGrad} and a DPM-based VC method similar to ours (Diff-VC) \cite{Popov2022} as baselines for comparison,
as these methods have been shown to perform as well as or better than many other state-of-the-art approaches for zero-shot any-to-any VC using non-parallel corpora.
Furthermore, in line with a recent trend in VC research that leverages the concept of neural audio codecs, we included FACodec\cite{Ju2024}, as implemented in NaturalSpeech 3 and representative of approaches that perform VC via discrete acoustic token sequences, as one of the baseline methods in our experiments.
To run Diff-VC and FACodec, we used the official source codes provided by the respective authors\footnote{Diff-VC: \url{https://github.com/trinhtuanvubk/Diff-VC}\\
FACodec: \url{https://github.com/lifeiteng/naturalspeech3_facodec}}.

\subsection{Model Setup}

The score network $\Vec{\epsilon}_{\theta}$ and vector field network $\Vec{v}_{\theta}$ were both implemented using the same architecture as VoiceGrad's score network \cite{Kameoka2024_VoiceGrad}, except that the number of output channels in each layer was set to 512 for VoiceGrad and 256 for LatentVoiceGrad.
The original VoiceGrad score network was a fully convolutional network with a U-Net structure consisting of 1D strided (transposed) convolution layers with gated linear units.
The only difference between the vector field network and the original score network lies in how the time parameter 
$t$ is incorporated into the network using sinusoidal positional encoding.
Specifically, the encoded vector was passed through three fully connected layers with two Mish activations in between, repeated along the time axis, and then added to the input of each convolution layer.
For the DPM version, the number of timesteps ($L$) was set to 20, the starting timestep ($L'$) in the reverse diffusion process was set to 18, and the noise variances $\{\sigma_l\}_{1\le l\le L}$ and $\{\beta_l\}_{1\le l \le L}$ were set as described in \cite{Kameoka2024_VoiceGrad}.

These networks 
were trained
using 
the Adam optimizer \cite{Kingma2015short} 
with random initialization,
the learning rate of 0.0002,
and 
the mini-batch size of 8.
All algorithms were implemented in PyTorch and run on a single Tesla V100 SXM2 GPU with a 32.0 GB memory and an Intel(R) Xeon(R) Gold 5218 16-core CPU @ 2.30GHz.
All versions of the proposed method were trained for 1000 epochs. The training time required for the score network was approximately 5 days for VoiceGrad and approximately 3 days for LatentVoiceGrad.
The peak GPU memory usage during the training was approximately 1,893 MB for VoiceGrad and 1,624 MB for LatentVoiceGrad.

\subsection{Objective Metrics}

We assessed the audio quality and intelligibility of the converted speech samples using predicted mean opinion scores (pMOS) and character error rate (CER) [\%] as objective metrics. The pMOS was obtained through Saeki's mean opinion score (MOS) predictor system \cite{Saeki2022}, which was submitted to the VoiceMOS challenge 2022 \cite{Huang2022} and demonstrated a strong correlation with human MOS ratings.
CER was measured using the wav2vec 2.0 model \cite{Baevski2020}, specifically the `Large LV-60K' architecture with an additional linear module, pre-trained on 60,000 hours of unlabeled audio from the Libri-Light \cite{Kahn2020} dataset and fine-tuned on 960 hours of transcribed audio from the LibriSpeech dataset \cite{Panayotov2015}.
To evaluate how likely each converted speech sample is to have been spoken by the target speaker,
we also measured the speaker embedding cosine similarity (SECS) between the embeddings from the converted and target speech samples (with a maximum value of $1$ and a minimum of $-1$). 
For computing SECS, we used publicly available source code provided online\footnote{https://github.com/resemble-ai/Resemblyzer}.
Higher pMOS and SECS scores, along with lower CER, indicate better performance.
To make visual comparison easier, the cells in the following tables are color-coded from red to green for first through fourth place.

\subsection{Choice of Input Feature Reresetnation}

Up to this point, our discussion has assumed that input speech is represented as a mel-spectrogram. However, in practice, alternative representations---such as short-time Fourier transform (STFT) amplitude spectrograms---can also be considered. Additionally, discrete acoustic token (DAT) sequences derived from a neural audio codec may serve as viable alternatives.
In this section, we evaluate the performance of VoiceGrad and LatentVoiceGrad using mel-spectrograms, STFT spectrograms, and DAT sequences as input features, and examine which representation is most suitable for our methods. 

For obtaining the DAT sequences, we chose to use the Descript Audio Codec (DAC) \cite{Kumar2023}, and employed the official source code provided by the authors\footnote{\url{https://github.com/descriptinc/descript-audio-codec?tab=readme-ov-file}}.
In our experiments, the dimensionality of the STFT spectrogram was 513, while that of the DAT representation was 1024. For all configurations, we used DPM as the underlying generative model.
When using STFT spectrograms, the output of the reverse diffusion process was converted to a mel-spectrogram before being passed to HiFi-GAN for waveform synthesis. In contrast, when DAT sequences were used, the converted DAT sequence was directly fed into the DAC decoder to generate the waveform.

Preliminary experiments revealed that in VoiceGrad, the score network architecture, which performed well with mel-spectrograms, was less effective for the STFT and DAT cases when used without modification. 
However, performance improved substantially when the number of output channels in each layer was doubled (from 512 to 1024). Based on this finding, we employed a score network with twice the number of output channels for the STFT and DAT cases compared to the mel-spectrogram case. This modification was applied only to VoiceGrad.
In contrast, LatentVoiceGrad used the same score network architecture across all input feature types. Instead, for the STFT and DAT cases, the autoencoder architecture was modified to double the number of output channels in each layer (from 256 to 512), while keeping the bottleneck dimensionality fixed at 32.
For LatentVoiceGrad, the autoencoder was trained with adversarial loss for all input types, following computation paths defined by their respective waveform generation processes. 

\reftabs{performance_feat}{rtf_feat} show the performance evaluation results and the real-time factors (RTFs) required for feature conversion, respectively, of VoiceGrad and LatentVoiceGrad when using mel-spectrograms, STFT spectrograms, and DAT sequences as input feature representations.
The experimental results revealed that, in VoiceGrad, which performs reverse diffusion directly on the input feature sequence, performance was highest when using mel-spectrograms as the feature representation, followed by STFT spectrograms and then DAT sequences. In particular, there was a noticeable performance gap between the STFT spectrogram and DAT cases. 
A similar trend was observed for LatentVoiceGrad; however, compared to VoiceGrad, it showed performance improvements in all cases---except for the DAT case, where only a slight improvement in pMOS was observed and a degradation in CER was noted.
This improvement is likely attributable to the compressed representations produced by the autoencoder, which make score function prediction easier than with the original high-dimensional input features.

Regarding conversion speed (RTF), VoiceGrad was fastest when using mel-spectrograms, followed by the STFT and DAT cases. This is primarily because the score network architecture was larger for the STFT and DAT cases than for the mel-spectrogram case. The difference in speed between the STFT and DAT cases can simply be attributed to the difference in feature dimensionality.
In contrast, LatentVoiceGrad achieved consistently high conversion speed across all input feature types. 
This is because the autoencoder's bottleneck dimensionality was fixed at a low value of 32, regardless of the input feature representation.

These results suggest that using mel-spectrograms as the input feature representation is a reasonable choice, considering both conversion quality and speed in the case of VoiceGrad, and conversion quality in the case of LatentVoiceGrad.
Overall, the DAT case underperformed compared to the mel-spectrogram and STFT cases.
However, this may be because optimal settings for the DAT case have not yet been identified, and it may be premature to conclude that DAT sequences are unsuitable as input features at this stage.

\begin{table}[t!]
\caption{Quality Comparison Across Input Feature Types}
\label{tab:performance_feat}
\centering
\begin{tabular}{l l c c c}
\thline
{\bf Method}&{\bf Feature}& {\bf pMOS}$\uparrow$ & {\bf CER}$\downarrow$ & {\bf SECS}$\uparrow$ \\\hline
\multirow{3}{*}{VoiceGrad}
&Mel & \cellcolor{color1}{3.86{\scriptsize $\pm$0.02}} & \cellcolor{color1}{2.88} & \cellcolor{color1}{0.830{\scriptsize $\pm$0.05}} \\
&STFT & \cellcolor{color3}{3.44{\scriptsize $\pm$0.03}} & \cellcolor{color3}{3.02} & \cellcolor{color3}{0.781{\scriptsize $\pm$0.05}} \\
&DAT & 2.41{\scriptsize $\pm$0.04} & 3.53 & 0.642{\scriptsize $\pm$0.07} \\
\hline
\multirow{3}{*}{LatentVoiceGrad}
&Mel & \cellcolor{color1}{3.93{\scriptsize $\pm$0.03}} & \cellcolor{color1}{2.99} & \cellcolor{color1}{0.844{\scriptsize $\pm$0.04}} \\
&STFT & \cellcolor{color3}{3.71{\scriptsize $\pm$0.02}} & \cellcolor{color3}{3.55} & \cellcolor{color3}{0.826{\scriptsize $\pm$0.05}} \\
&DAT & 2.43{\scriptsize $\pm$0.03} & {9.08} & 0.793{\scriptsize $\pm$0.05} \\
\thline
\end{tabular}
\end{table}

\begin{table}[t!]
\caption{RTF Comparison Across Input Feature Types}
\label{tab:rtf_feat}
\centering
\begin{tabular}{l l c c}
\thline
\multirow{2}{*}{{\bf Method}}&\multirow{2}{*}{{\bf Feature}} & \multicolumn{2}{c}{{\bf RTF}$\downarrow$} \\
&&GPU&CPU\\\hline
\multirow{3}{*}{VoiceGrad}
&Mel& 0.045& 0.227\\
&STFT& 0.224& 0.852\\
&DAT& 0.248& 1.490 \\
\hline
\multirow{3}{*}{LatentVoiceGrad}
&Mel & 0.034 & 0.166\\
&STFT & 0.040 & 0.179 \\
&DAT & 0.038 & 0.175 \\
\thline
\end{tabular}
\end{table}

\subsection{Impact of Adversarial Training in LatentVoiceGrad}

\begin{table}[t!]
\caption{Baseline Comparison}
\label{tab:baseline_comp}
\centering
\begin{tabular}{l c c c}
\thline
{\bf Method}& {\bf pMOS}$\uparrow$ & {\bf CER}$\downarrow$ & {\bf SECS}$\uparrow$ \\\hline
Ground truth & 4.13{\scriptsize $\pm$0.04} & 1.42 & 1.0{\scriptsize $\pm$0.0} \\
Diff-VC \cite{Popov2022} 
& 3.67{\scriptsize $\pm$0.02} & 8.20 & 0.813{\scriptsize $\pm$0.05} \\
FACodec \cite{Ju2024} 
& 3.71{\scriptsize $\pm$0.03} & 2.56 & 0.808{\scriptsize $\pm$0.06} \\
VoiceGrad-DPM \cite{Kameoka2024_VoiceGrad} 
& \cellcolor{color3}{3.86{\scriptsize $\pm$0.02}} & \cellcolor{color1}{2.88} & \cellcolor{color2}{0.830{\scriptsize $\pm$0.05}} \\
LatentVoiceGrad-DPM 
& \cellcolor{color1}{3.93{\scriptsize $\pm$0.01}} & \cellcolor{color3}{2.99} & \cellcolor{color1}{0.844{\scriptsize $\pm$0.04}} \\
VoiceGrad-FM 
& \cellcolor{color4}{3.85{\scriptsize $\pm$0.02}} & \cellcolor{color2}{2.95} & \cellcolor{color4}{0.817{\scriptsize $\pm$0.05}} \\
LatentVoiceGrad-FM & \cellcolor{color2}{3.92{\scriptsize $\pm$0.02}} & \cellcolor{color4}{3.21} & \cellcolor{color3}{0.827{\scriptsize $\pm$0.05}} \\
\thline
\end{tabular}
\end{table}

To evaluate the impact of including adversarial loss in training the autoencoder for LatentVoiceGrad, as outlined in \refsubsec{aegan}, we compared the quality of LatentVoiceGrad's converted speech samples under two conditions: when the autoencoder was trained using only the reconstruction loss $\mathcal{J}_{\rm rec}$ and KL loss $\mathcal{J}_{\rm KL}$, and when it was trained using the full objective.
The results, shown in \reftab{effect_adversarial}, refer to these conditions as `Regular' for the 
$\mathcal{J}_{\rm rec} + \mathcal{J}_{\rm KL}$ objective and `Adversarial' for the full objective.
For this comparison, the underlying generative model used was DPM.
As shown in \reftab{effect_adversarial}, the autoencoder trained with the full objective that includes adversarial loss demonstrated significantly better conversion performance in terms of audio quality and speaker similarity.

\begin{table}[t!]
\caption{Effect of Adversarial Autoencoder Training}
\label{tab:effect_adversarial}
\centering
\begin{tabular}{l c c c}
\thline
{\bf AE Training}& {\bf pMOS}$\uparrow$ & {\bf CER}$\downarrow$ & {\bf SECS}$\uparrow$ \\\hline
Regular & 3.78{\scriptsize $\pm$0.02} & \cellcolor{color1}{2.95} & 0.829{\scriptsize $\pm$0.05} \\
Adversarial & \cellcolor{color1}{3.93{\scriptsize $\pm$0.01}} & \cellcolor{color5}{2.99} & \cellcolor{color1}{0.844{\scriptsize $\pm$0.04}} \\
\thline
\end{tabular}
\end{table}

\subsection{Impact of $r$ and $L$ Settings in FM Versions}
In the FM versions, the fraction $r$ of noise added to the initial point and the number of iterations $L$ can be set arbitrarily. We evaluated how the performance changes with different settings of these parameters: \reftab{effect_r} shows the results for different values of $r$ when $L=10$, and \reftab{effect_L} presents the results for different values of $L$ when $r=0.7$.

The results in \reftab{effect_r} indicate that the quality of converted speech is better when $r$ is around 0.3 to 0.7 for VoiceGrad-FM and around 0.5 to 0.7 for LatentVoiceGrad-FM. This suggests that for both methods, it is more effective to start the ODE with the feature sequence of the source speech mixed with a moderate amount of noise. However, the CER results indicate that as more noise is added to the initial point, the intelligibility of the converted speech decreases. This suggests that using the feature sequence of the source speech as is may be preferable from a CER perspective, as retaining more information from the source speech helps preserve linguistic features. On the other hand, this can reduce SECS, leading to a decrease in similarity to the target speaker. 
A similar effect was observed in the DPM version of VoiceGrad, where it was found experimentally that adding a moderate amount of noise to the initial point of the reverse diffusion process produced better results. Based on these findings, setting $r$ to about 0.7 seems to achieve a good balance between audio quality, intelligibility, and speaker similarity.

\begin{table}[t!]
\caption{Impact of $r$ Setting in FM Versions}
\label{tab:effect_r}
\centering
\begin{tabular}{l c c c c}
\thline
{\bf Method}&{\bf $r$}& {\bf pMOS}$\uparrow$ & {\bf CER}$\downarrow$ & {\bf SECS}$\uparrow$ \\\hline
\multirow{11}{*}{VoiceGrad-FM}
&0.0 & 3.76{\scriptsize $\pm$0.02} & \cellcolor{color2}{2.79} & 0.746{\scriptsize $\pm$0.05} \\
&0.1 & 3.82{\scriptsize $\pm$0.02} & \cellcolor{color1}{2.78} & 0.757{\scriptsize $\pm$0.05} \\
&0.2 & 3.87{\scriptsize $\pm$0.02} & \cellcolor{color3}{2.86} & 0.767{\scriptsize $\pm$0.05} \\
&0.3 & \cellcolor{color2}{3.90{\scriptsize $\pm$0.02}} & \cellcolor{color4}{2.91} & 0.778{\scriptsize $\pm$0.05} \\
&0.4 & \cellcolor{color1}{3.91{\scriptsize $\pm$0.02}} & 2.92 & 0.788{\scriptsize $\pm$0.05} \\
&0.5 & \cellcolor{color2}{3.90{\scriptsize $\pm$0.02}} & 2.98 & 0.799{\scriptsize $\pm$0.05} \\
&0.6 & \cellcolor{color4}{3.89{\scriptsize $\pm$0.02}} & 2.97 & 0.807{\scriptsize $\pm$0.05} \\
&0.7 & 3.85{\scriptsize $\pm$0.02} & 2.95 & \cellcolor{color4}{0.817{\scriptsize $\pm$0.05}} \\
&0.8 & 3.79{\scriptsize $\pm$0.02} & 3.15 & \cellcolor{color3}{0.823{\scriptsize $\pm$0.05}} \\
&0.9 & 3.74{\scriptsize $\pm$0.02} & 3.11 & \cellcolor{color2}{0.829{\scriptsize $\pm$0.05}} \\
&1.0 & 3.65{\scriptsize $\pm$0.02} & 3.39 & \cellcolor{color1}{0.834{\scriptsize $\pm$0.05}} \\\hline
\multirow{11}{*}{LatentVoiceGrad-FM}
&0.0 & 3.27{\scriptsize $\pm$0.03} & 3.12 & 0.728{\scriptsize $\pm$0.05} \\
&0.1 & 3.44{\scriptsize $\pm$0.03} & \cellcolor{color4}{2.98} & 0.747{\scriptsize $\pm$0.05} \\
&0.2 & 3.60{\scriptsize $\pm$0.03} & \cellcolor{color1}{2.94} & 0.764{\scriptsize $\pm$0.05} \\
&0.3 & 3.73{\scriptsize $\pm$0.02} & \cellcolor{color1}{2.94} & 0.781{\scriptsize $\pm$0.05} \\
&0.4 & 3.83{\scriptsize $\pm$0.02} & \cellcolor{color3}{2.97} & 0.796{\scriptsize $\pm$0.05} \\
&0.5 & \cellcolor{color3}{3.89{\scriptsize $\pm$0.02}} & 3.02 & 0.808{\scriptsize $\pm$0.05} \\
&0.6 & \cellcolor{color1}{3.93{\scriptsize $\pm$0.02}} & 3.15 & 0.819{\scriptsize $\pm$0.05} \\
&0.7 & \cellcolor{color2}{3.92{\scriptsize $\pm$0.02}} & 3.21 & \cellcolor{color4}{0.827{\scriptsize $\pm$0.05}} \\
&0.8 & \cellcolor{color4}{3.88{\scriptsize $\pm$0.02}} & 3.21 & \cellcolor{color3}{0.833{\scriptsize $\pm$0.05}} \\
&0.9 & 3.85{\scriptsize $\pm$0.02} & 3.34 & \cellcolor{color2}{0.838{\scriptsize $\pm$0.05}} \\
&1.0 & 3.77{\scriptsize $\pm$0.02} & 3.60 & \cellcolor{color1}{0.840{\scriptsize $\pm$0.05}} \\
\thline
\end{tabular}
\end{table}

In general, with Euler's method, increasing the number of iterations improves the accuracy of the ODE solution. Therefore, it is crucial to understand how the quality of the converted speech varies with different settings of $L$. The results in \reftab{effect_L} show that, as expected, increasing $L$ improves performance. However, it is noteworthy that even with just 3 iterations, the performance remains relatively high. This suggests that the FM version can deliver good results with a small number of iterations, which helps reduce processing time.
Since the SECS increases steadily with higher $L$ values, it may be beneficial to use a moderate value like 10 in scenarios where speaker similarity is particularly important. The CER results also confirm that the intelligibility of the converted speech is well maintained when $L$ is greater than 2.

\begin{table}[t!]
\caption{Performance w.r.t. $L$ in FM Versions}
\label{tab:effect_L}
\centering
\begin{tabular}{l c c c c}
\thline
{\bf Method}&{\bf $L$}& {\bf pMOS}$\uparrow$ & {\bf CER}$\downarrow$ & {\bf SECS}$\uparrow$ \\\hline
\multirow{6}{*}{VoiceGrad-FM}
&1 & 1.24{\scriptsize $\pm$0.00} & 33.50 & 0.570{\scriptsize $\pm$0.06} \\
&2 & 3.64{\scriptsize $\pm$0.01} & \cellcolor{color2}{3.01} & 0.751{\scriptsize $\pm$0.05} \\
&3 & \cellcolor{color2}{3.85{\scriptsize $\pm$0.02}} & \cellcolor{color4}{3.06} & \cellcolor{color3}{0.783{\scriptsize $\pm$0.05}} \\
&5 & \cellcolor{color1}{3.89{\scriptsize $\pm$0.02}} & 3.11 & \cellcolor{color4}{0.804{\scriptsize $\pm$0.05}} \\
&10 & \cellcolor{color2}{3.85{\scriptsize $\pm$0.01}} & \cellcolor{color1}{2.95} & \cellcolor{color2}{0.817{\scriptsize $\pm$0.05}} \\
&20 & \cellcolor{color4}{3.81{\scriptsize $\pm$0.02}} & \cellcolor{color3}{3.03} & \cellcolor{color1}{0.820{\scriptsize $\pm$0.05}} \\
\hline
\multirow{6}{*}{LatentVoiceGrad-FM}
&1 & 1.24{\scriptsize $\pm$0.00} & 36.11 & 0.625{\scriptsize $\pm$0.06} \\
&2 & 3.26{\scriptsize $\pm$0.02} & 3.42 & 0.751{\scriptsize $\pm$0.05} \\
&3 & \cellcolor{color4}{3.75{\scriptsize $\pm$0.02}} & \cellcolor{color4}{3.28} & \cellcolor{color4}{0.784{\scriptsize $\pm$0.05}} \\
&5 & \cellcolor{color1}{3.93{\scriptsize $\pm$0.02}} & \cellcolor{color3}{3.24} & \cellcolor{color3}{0.811{\scriptsize $\pm$0.05}} \\
&10 & \cellcolor{color1}{3.93{\scriptsize $\pm$0.02}} & \cellcolor{color2}{3.21} & \cellcolor{color2}{0.827{\scriptsize $\pm$0.05}} \\
&20 & \cellcolor{color3}{3.86{\scriptsize $\pm$0.02}} & \cellcolor{color1}{3.17} & \cellcolor{color1}{0.833{\scriptsize $\pm$0.05}} \\
\thline
\end{tabular}
\end{table}

\subsection{Baseline Comparison}

\reftab{baseline_comp} presents the results for Diff-VC, FACodec, and the four versions of VoiceGrad, alongside metrics for the ground truth speech. As the data shows, all four VoiceGrad versions significantly outperformed Diff-VC and FACodec. 
Among the VoiceGrad versions, LatentVoiceGrad-DPM and LatentVoiceGrad-FM excelled in pMOS, while VoiceGrad-DPM and VoiceGrad-FM achieved better CER results. LatentVoiceGrad-DPM and VoiceGrad-DPM also surpassed the other methods in SECS. These results indicate that the latent versions are superior in audio quality, and the DPM versions excel in intelligibility and speaker similarity. Moreover, the FM versions offer the advantage of faster conversion speeds, which will be further discussed in the next section. Each version has distinct advantages, so it is important to select the appropriate version based on the specific requirements and scenarios. 

\begin{table}[t!]
\caption{Performance in Cross-Dataset Scenario}
\label{tab:crossdataset}
\centering
\begin{tabular}{l c c c}
\thline
{\bf Method}& {\bf pMOS}$\uparrow$ & {\bf CER}$\downarrow$ & {\bf SECS}$\uparrow$ \\\hline
Ground truth & 4.04{\scriptsize $\pm$0.01} & 1.42 & 1.0{\scriptsize $\pm$0.0} \\
Diff-VC \cite{Popov2022} 
& 3.60{\scriptsize $\pm$0.01} & 8.75 & \cellcolor{color1}{0.809{\scriptsize $\pm$0.07}} \\
FACodec \cite{Ju2024} 
& 3.57{\scriptsize $\pm$0.01} & \cellcolor{color1}{2.52} & 0.734{\scriptsize $\pm$0.08} \\
VoiceGrad-DPM \cite{Kameoka2024_VoiceGrad} 
& \cellcolor{color2}{3.92{\scriptsize $\pm$0.01}} & \cellcolor{color3}{2.90} & \cellcolor{color3}{0.755{\scriptsize $\pm$0.07}} \\
LatentVoiceGrad-DPM 
& \cellcolor{color1}{3.96{\scriptsize $\pm$0.01}} & \cellcolor{color4}{3.04} & \cellcolor{color2}{0.756{\scriptsize $\pm$0.07}} \\
VoiceGrad-FM 
& \cellcolor{color4}{3.88{\scriptsize $\pm$0.01}} & \cellcolor{color2}{2.89} & \cellcolor{color4}{0.754{\scriptsize $\pm$0.07}} \\
LatentVoiceGrad-FM & \cellcolor{color3}{3.91{\scriptsize $\pm$0.01}} & \cellcolor{color5}{3.20} & \cellcolor{color5}{0.752{\scriptsize $\pm$0.06}} \\
\thline
\end{tabular}
\end{table}

\subsection{Performance in Cross-Dataset Scenario}

In the previous experiments, we evaluated VC performance in an {\it unseen-to-unseen} scenario, where neither the source nor the target speaker's speech was included in the training data. To further assess the generalizability of each method, we also conducted experiments in a {\it cross-dataset} scenario, where the source and target speakers were not only excluded from the training data but also came from different datasets. For this experiment, we selected the 39 speakers from the ``test-clean'' subset of the LibriTTS dataset \cite{Zen2019} as the target speakers. The pMOS, CER, and SECS results for each method under this condition are presented in \reftab{crossdataset}.
According to these results, while all methods showed a slight decrease in SECS compared to the condition where source and target speakers were drawn from the same dataset, the scores remained reasonably high. 
As for pMOS and CER, while Diff-VC and FACodec exhibited a slight degradation in performance, both VoiceGrad and LatentVoiceGrad maintained nearly the same level of performance as in the same-dataset condition. These findings suggest that the performance of our methods is not strongly dependent on the dataset used.

\subsection{Real-Time Factor for Mel-Spectrogram Conversion}

\reftab{rtf} shows the real-time factors (RTFs) required for mel-spectrogram conversion by each method, 
measured on both GPU and CPU. 
Here, $L$ denotes the number of iterations in the reverse diffusion process or the ODE solving process.
For LatentVoiceGrad, the processing time includes the time required to encode and decode mel-spectrograms using the autoencoder. 
Similarly, for Diff-VC, the processing time accounts for encoding the source speech mel-spectrograms into average voice mel-spectrograms. 
Given that the original Diff-VC paper examined cases with $L=6$ and $L=30$, we measured the RTFs for both scenarios here, labeled as `Diff-VC-6' and `Diff-VC-30', respectively.
Note that all of these values do not include the time taken for waveform generation by the neural vocoder (HiFi-GAN).
Also, since FACodec performs VC without involving mel-spectrograms in the conversion process, its end-to-end real-time factor was measured solely for reference purposes.

\begin{table}[t!]
\caption{
Real-time factors required for mel-spectrogram conversion (end-to-end conversion in the case of FACodec only)
}
\label{tab:rtf}
\centering
\begin{tabular}{l c c c}
\thline
\multirow{2}{*}{{\bf Method}}&\multirow{2}{*}{{\bf $L$}} & \multicolumn{2}{c}{{\bf RTF}$\downarrow$} \\
&&GPU&CPU\\\hline
\multirow{2}{*}{Diff-VC}
&6 & 0.130 & 0.968\\
&30 & 0.535& 4.495\\
\hline
\multirow{1}{*}{FACodec}
&--- & 0.278 & 0.436\\
\hline
\multirow{1}{*}{VoiceGrad-DPM}
&20 & 0.045 & 0.227\\
\hline
\multirow{1}{*}{LatentVoiceGrad-DPM}
&20 & 0.034 &0.166\\
\hline
\multirow{6}{*}{VoiceGrad-FM}
&1 & 0.003 &0.012\\
&2 & 0.005 &0.024\\
&3 & 0.007 &0.035\\
&5 & 0.012 &0.060\\
&10 & 0.023 &0.114\\
&20 & 0.045 &0.229\\
\hline
\multirow{6}{*}{LatentVoiceGrad-FM}
&1 & 0.008 &0.064\\
&2 & 0.010 &0.069\\
&3 & 0.011 &0.075\\
&5 & 0.014 &0.085\\
&10 & 0.022 &0.111\\
&20 & 0.036 &0.164\\
\thline
\end{tabular}
\end{table}

The results show that VoiceGrad-DPM with $L=20$ was approximately 12 times faster than Diff-VC-30 and 3 times faster than Diff-VC-6, likely due to the simpler architecture of the score network used in VoiceGrad compared to Diff-VC. When comparing VoiceGrad-DPM and LatentVoiceGrad-DPM, we found that LatentVoiceGrad-DPM was slightly faster, even with the additional processing required by the autoencoder. This is because the autoencoder's bottleneck features are of lower dimension than mel-spectra, allowing for a slightly smaller score network. Comparing VoiceGrad-DPM and VoiceGrad-FM, the FM version benefits from flexible iteration settings during testing, so its processing time varies depending on the number of iterations. When the iterations match those of the DPM version, processing times were nearly identical due to the shared architecture of the score and vector field networks. However, with fewer iterations, the FM version processed naturally faster. 
Lastly, when comparing VoiceGrad-FM and LatentVoiceGrad-FM, the latter was more efficient with a high number of iterations but less so with fewer iterations. This is because LatentVoiceGrad-FM requires additional processing by the autoencoder, which does not depend on the number of iterations. When $L=10$, the processing times for both methods were nearly identical.

LatentVoiceGrad-DPM and LatentVoiceGrad-FM (with $L=10$) offer conversion quality that is comparable to or better than VoiceGrad-DPM, with processing times at 70\% and 50\% of those for VoiceGrad-DPM, respectively. 
These results highlight the effectiveness of the proposed improvements made to VoiceGrad.
Generally, the latent version primarily improved conversion performance, while the FM version contributed to increasing processing speed.

\subsection{Subjective Evaluation}

\begin{table}[t!]
\caption{Results of the Listening Tests}
\vspace{-2ex}
\centering
\begin{tabular}{l V{3} c | c }
\thline
{\bf Methods}&{\bf qMOS}$\uparrow$&{\bf sMOS}$\uparrow$
\\\hline
Ground truth& 4.21{\scriptsize $\pm$0.19}&--- \\
Diff-VC&3.56{\scriptsize $\pm$0.13}& 2.74{\scriptsize $\pm$0.20}\\
VoiceGrad-DPM&3.83{\scriptsize $\pm$0.07}&2.63{\scriptsize $\pm$0.14}\\
LatentVoiceGrad-DPM&4.09{\scriptsize $\pm$0.08}&3.05{\scriptsize $\pm$0.15}\\
VoiceGrad-FM&4.07{\scriptsize $\pm$0.07}&2.74{\scriptsize $\pm$0.14}\\
LatentVoiceGrad-FM&4.16{\scriptsize $\pm$0.09}&3.02{\scriptsize $\pm$0.17}\\
\thline
\end{tabular}
\label{tab:MOS}
\end{table}

We conducted MOS tests to compare the audio quality and speaker similarity of the converted speech samples generated by the proposed and baseline methods. Twelve listeners participated in both tests, which were conducted online. Each participant was asked to use headphones in a quiet environment. For the audio quality test, we included ground truth real speech samples from the target speakers as references. Each listener rated the naturalness of each sample on a five-point scale: 5 for Excellent, 4 for Good, 3 for Fair, 2 for Poor, and 1 for Bad. These scores will be referred to as ``qMOS''.
For the speaker similarity test, each listener was given pairs of converted speech samples and real speech sample from the corresponding target speakers. They were asked to evaluate how likely the two samples were produced by the same speaker, using a four-point scale: 4 for Same (sure), 3 for Same (not sure), 2 for Different (not sure), and 1 for Different (sure). These scores will be referred to as ``sMOS''. 
The qMOS and sMOS scores, along with their 95\% confidence intervals, are shown in \reftab{MOS}.
The results indicate that the latent versions (i.e., LatentVoiceGrad-DPM and LatentVoiceGrad-FM) outperformed the regular mel-spectrogram-domain versions (VoiceGrad-DPM and VoiceGrad-FM) in both qMOS and sMOS scores. Furthermore, switching from DPM to FM as the underlying generative model in both the latent and regular versions also led to improvements in these scores.
Notably, the converted samples generated by LatentVoiceGrad-FM were very close in audio quality to the real speech samples.
The results of the current experiment indicate that the original VoiceGrad outperforms Diff-VC in audio quality, while Diff-VC surpasses VoiceGrad in speaker similarity. However, by incorporating one or both of the ideas from the latent and FM versions, it is possible to achieve better performance than either of these two methods in both audio quality and speaker similarity.
Consistent with the results of the earlier objective evaluation, these findings demonstrate the effectiveness of the two improvements made to the original VoiceGrad.

\begin{figure*}[t!]
\centering
\begin{minipage}[t!]{.245\linewidth}
  \centerline{\includegraphics[width=.98\linewidth]{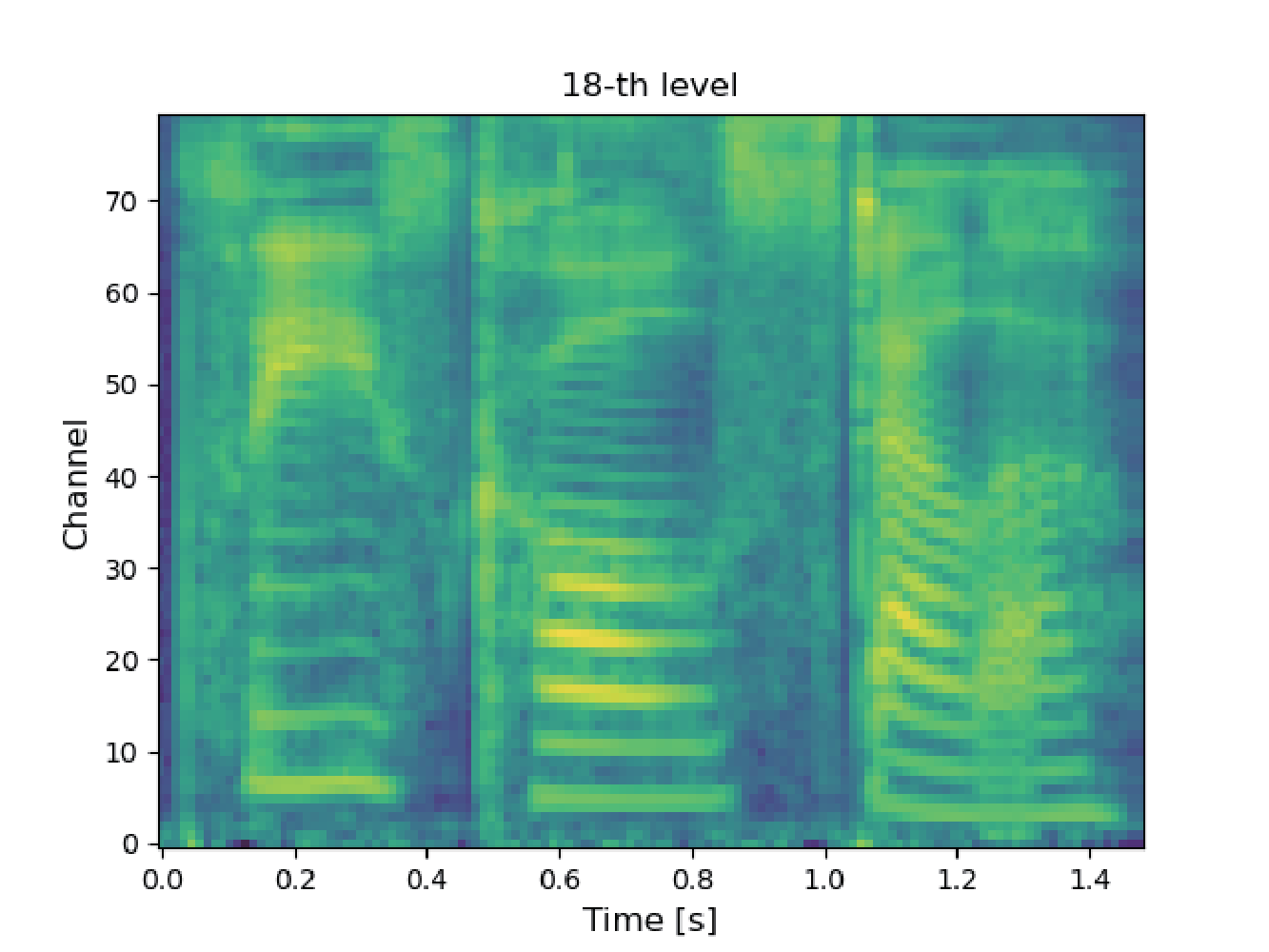}}
\end{minipage}
\begin{minipage}[t!]{.245\linewidth}
    \centerline{\includegraphics[width=.98\linewidth]{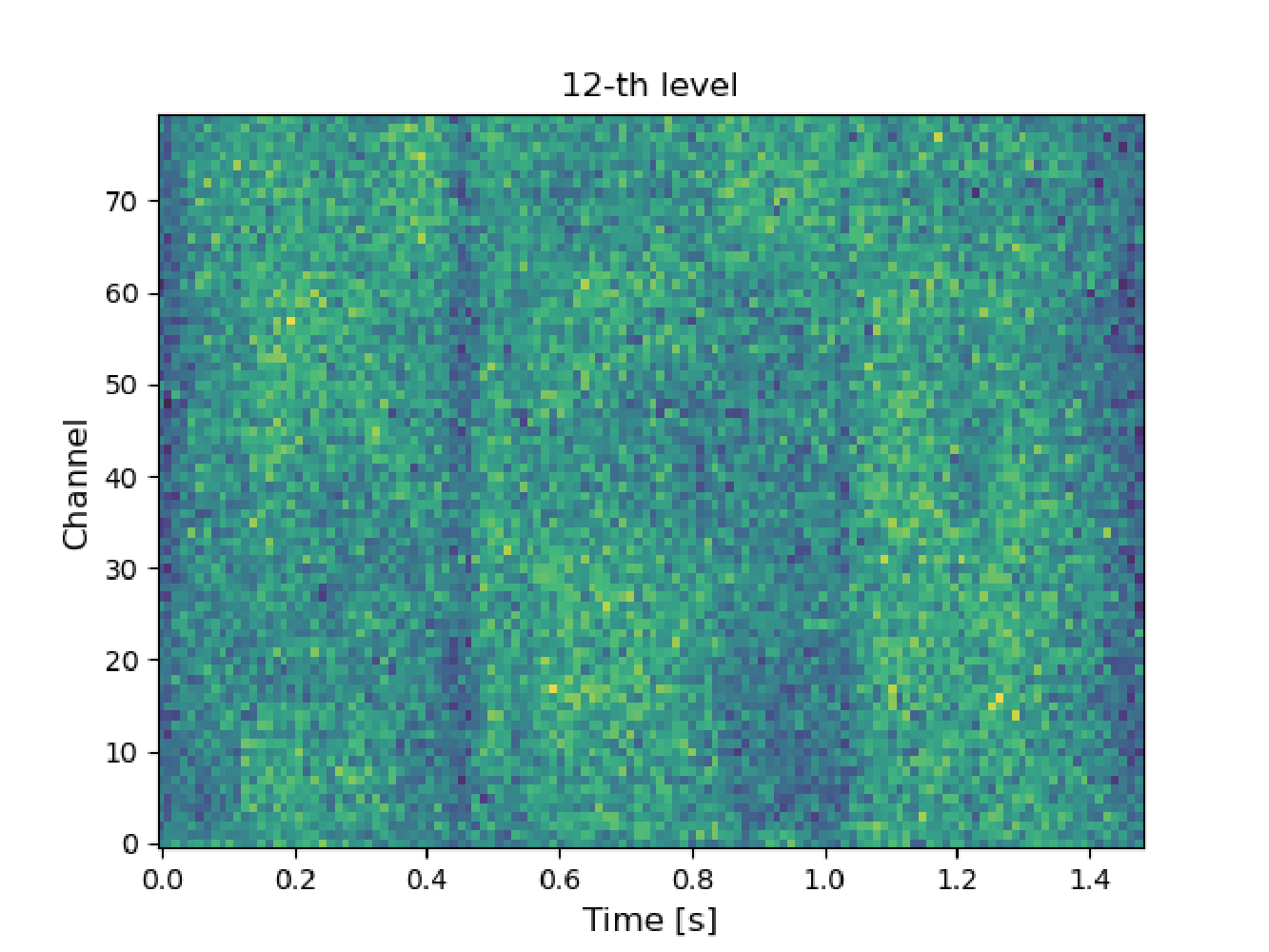}}
    \end{minipage}
\begin{minipage}[t!]{.245\linewidth}
      \centerline{\includegraphics[width=.98\linewidth]{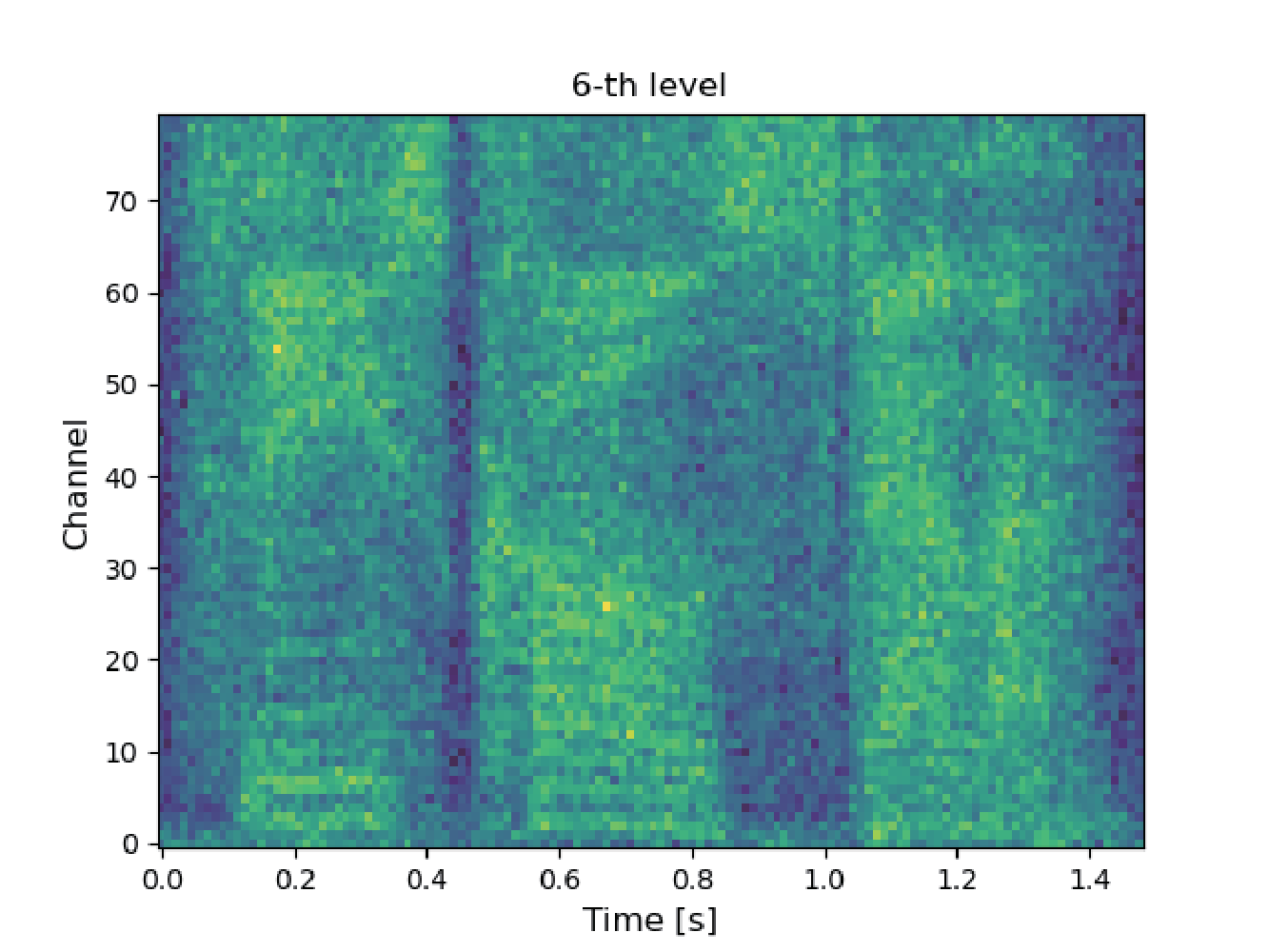}}
      \end{minipage}
\begin{minipage}[t!]{.245\linewidth}
        \centerline{\includegraphics[width=.98\linewidth]{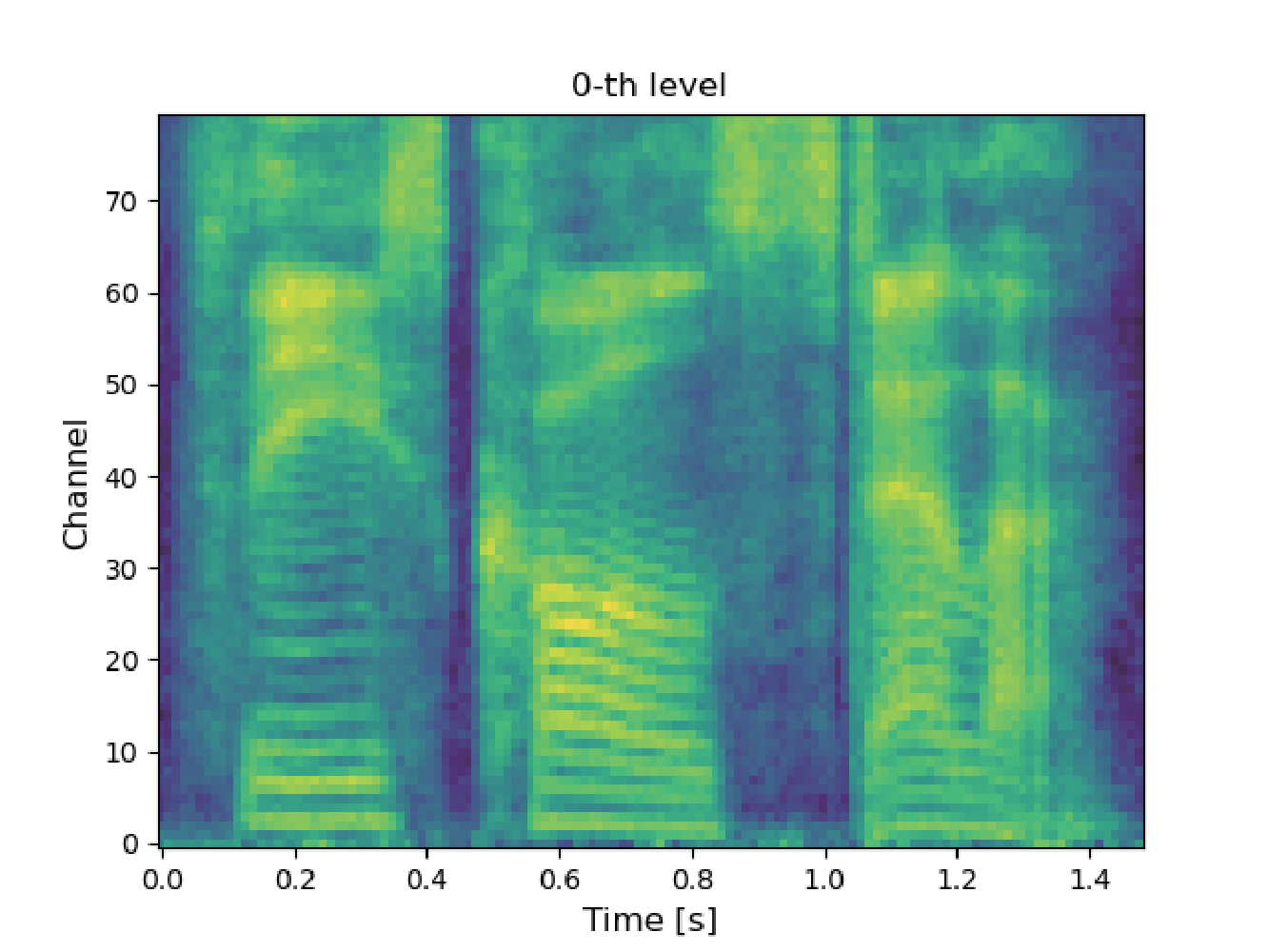}}
        \end{minipage}
 \vspace{-0ex}
  \centering
  \caption{Mel-spectrogram updates at steps 18, 12, 6, and 0 during VoiceGrad-DPM's VC process, using speaker p238 (female)'s utterance as input and speaker p241 (male)'s utterance as the reference speech (from left to right).}
\label{fig:voicegrad-dmp_vc-process}
\end{figure*}

Finally, examples of the mel-spectrograms and autoencoder bottleneck feature sequences at each step of the VoiceGrad-DPM, LatentVoiceGrad-DPM, VoiceGrad-FM, and LatentVoiceGrad-FM VC processes are shown in \reffigss{voicegrad-dmp_vc-process}{latentvoicegrad-fm_vc-process}, respectively.
Note that in LatentVoiceGrad, although the mel-spectrogram is not computed at every step during the actual iterative process, here it is decoded from the bottleneck feature sequence for visualization purposes.
Also, note that the DPM version progresses in descending step order, while the FM version progresses in ascending step order.
These graphs reveal that the mel-spectrogram changes differently for each approach, reflecting the specific space being updated and the planned update path.

Lastly, samples of the converted speech generated by the proposed methods can be found on our website\footnote{\url{https://www.kecl.ntt.co.jp/people/kameoka.hirokazu/Demos/latentvoicegrad/index.html}}.

\begin{figure*}[t!]
\centering
\begin{minipage}[t!]{.245\linewidth}
  \centerline{\includegraphics[width=.98\linewidth]{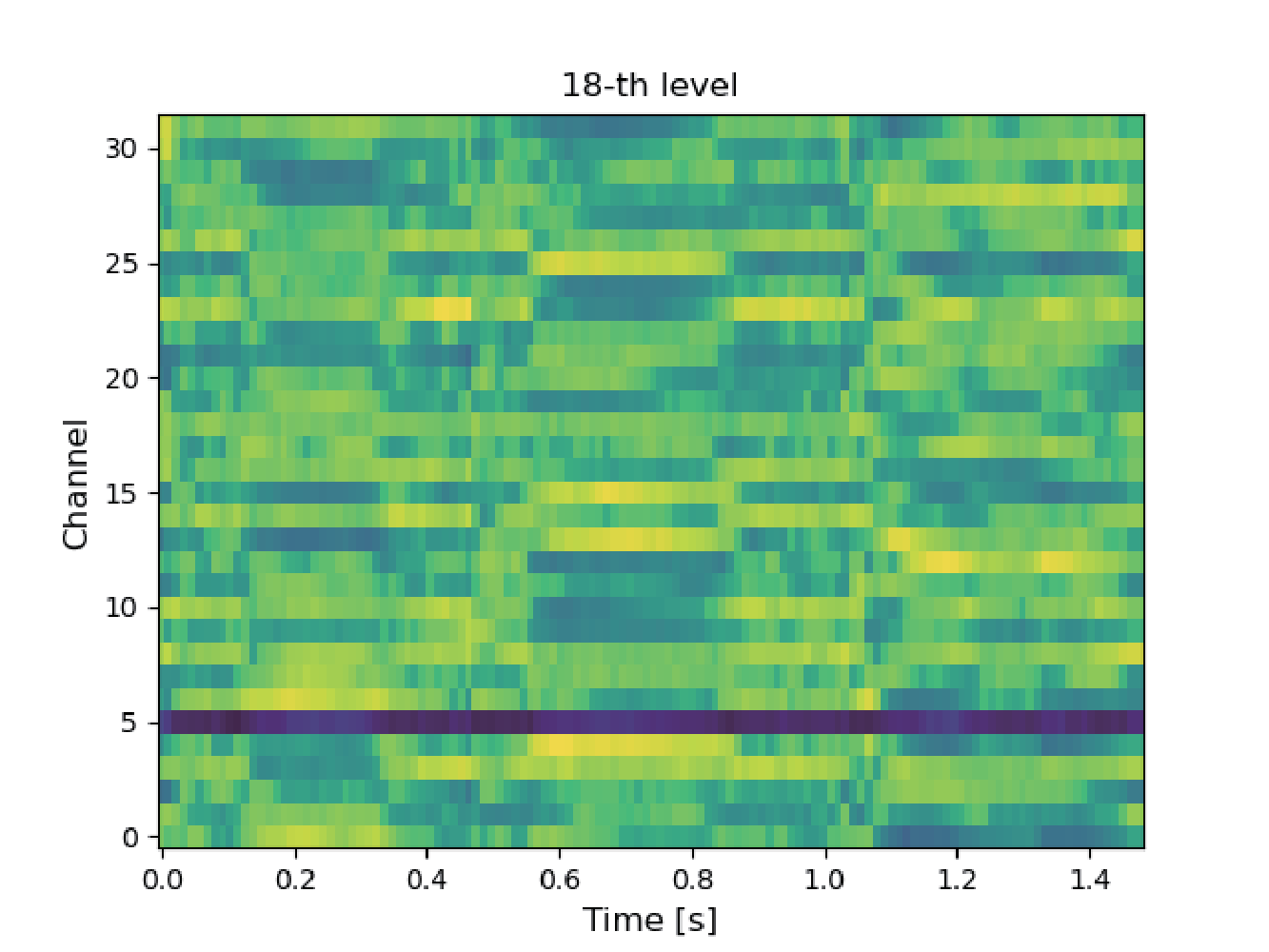}}
  \centerline{\includegraphics[width=.98\linewidth]{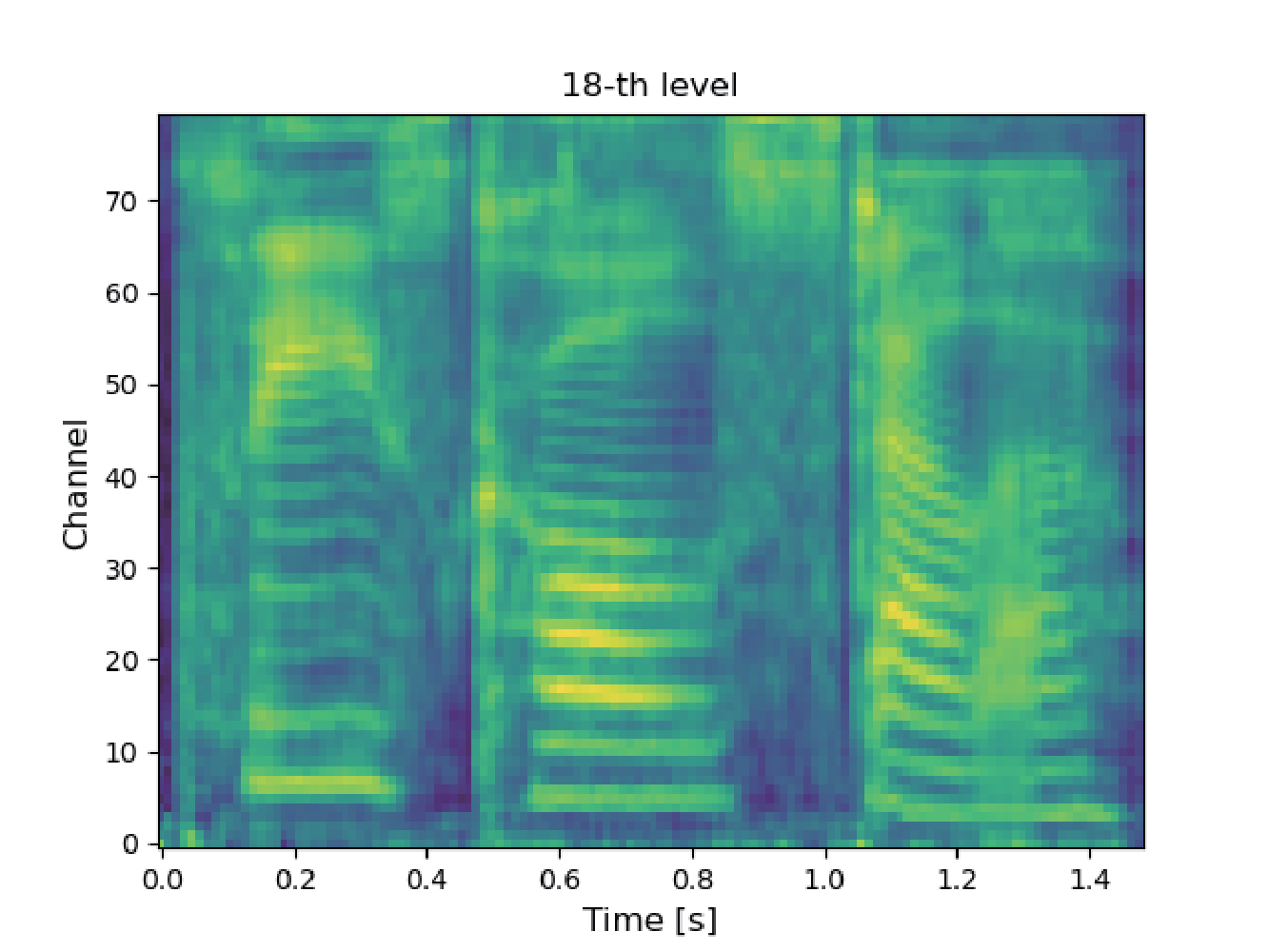}}
\end{minipage}
\begin{minipage}[t!]{.245\linewidth}
    \centerline{\includegraphics[width=.98\linewidth]{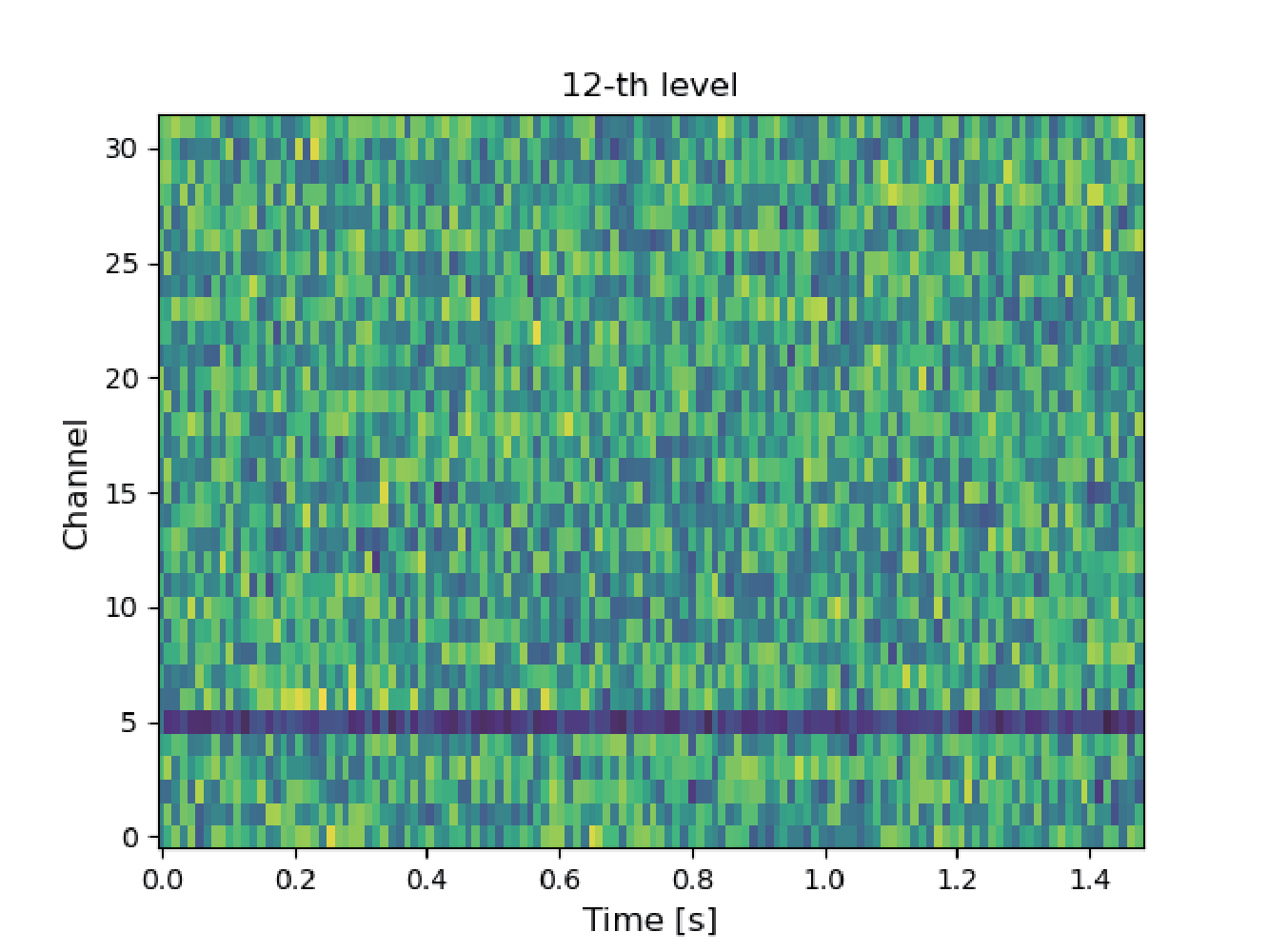}}
    \centerline{\includegraphics[width=.98\linewidth]{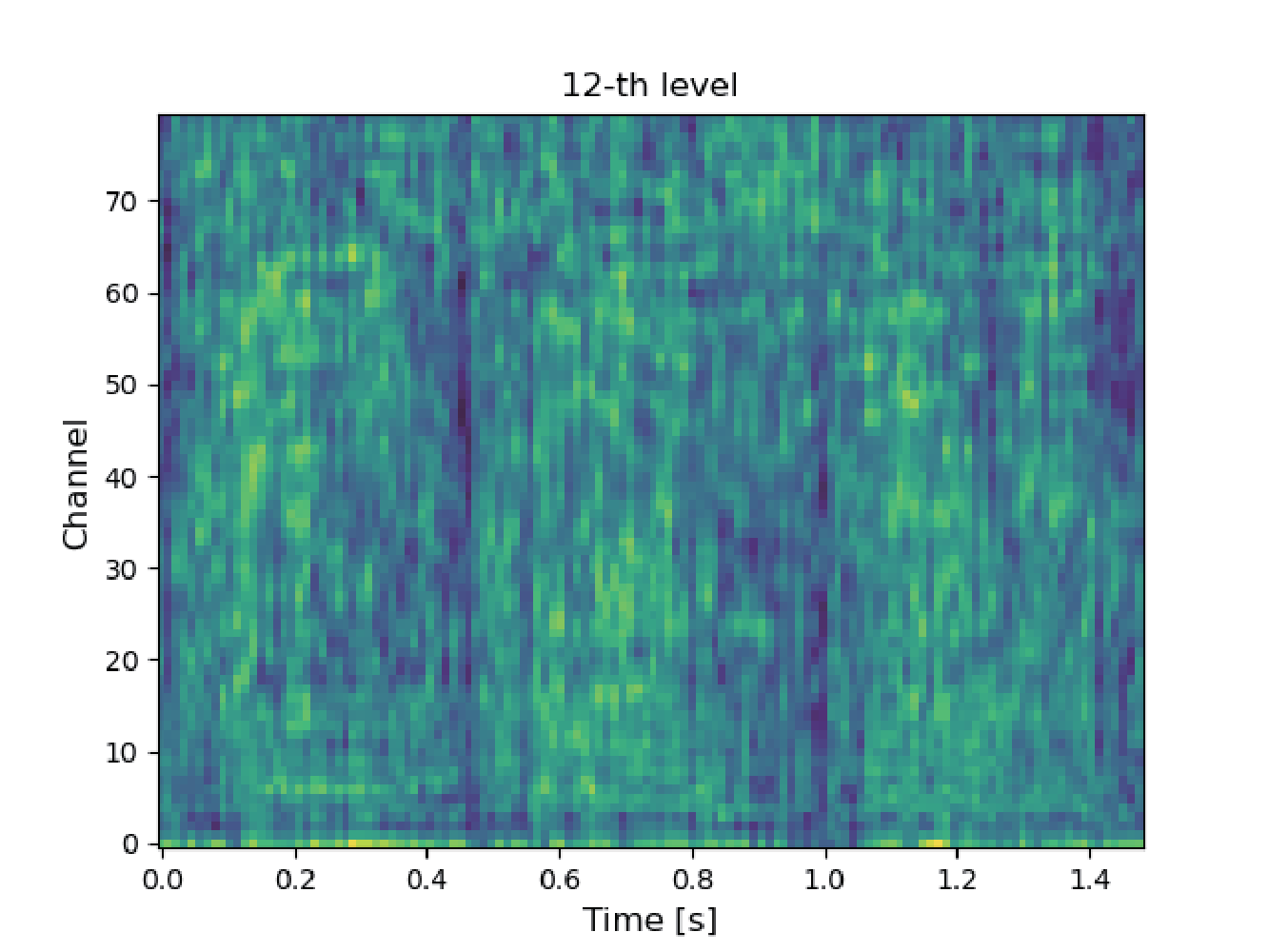}}
    \end{minipage}
\begin{minipage}[t!]{.245\linewidth}
      \centerline{\includegraphics[width=.98\linewidth]{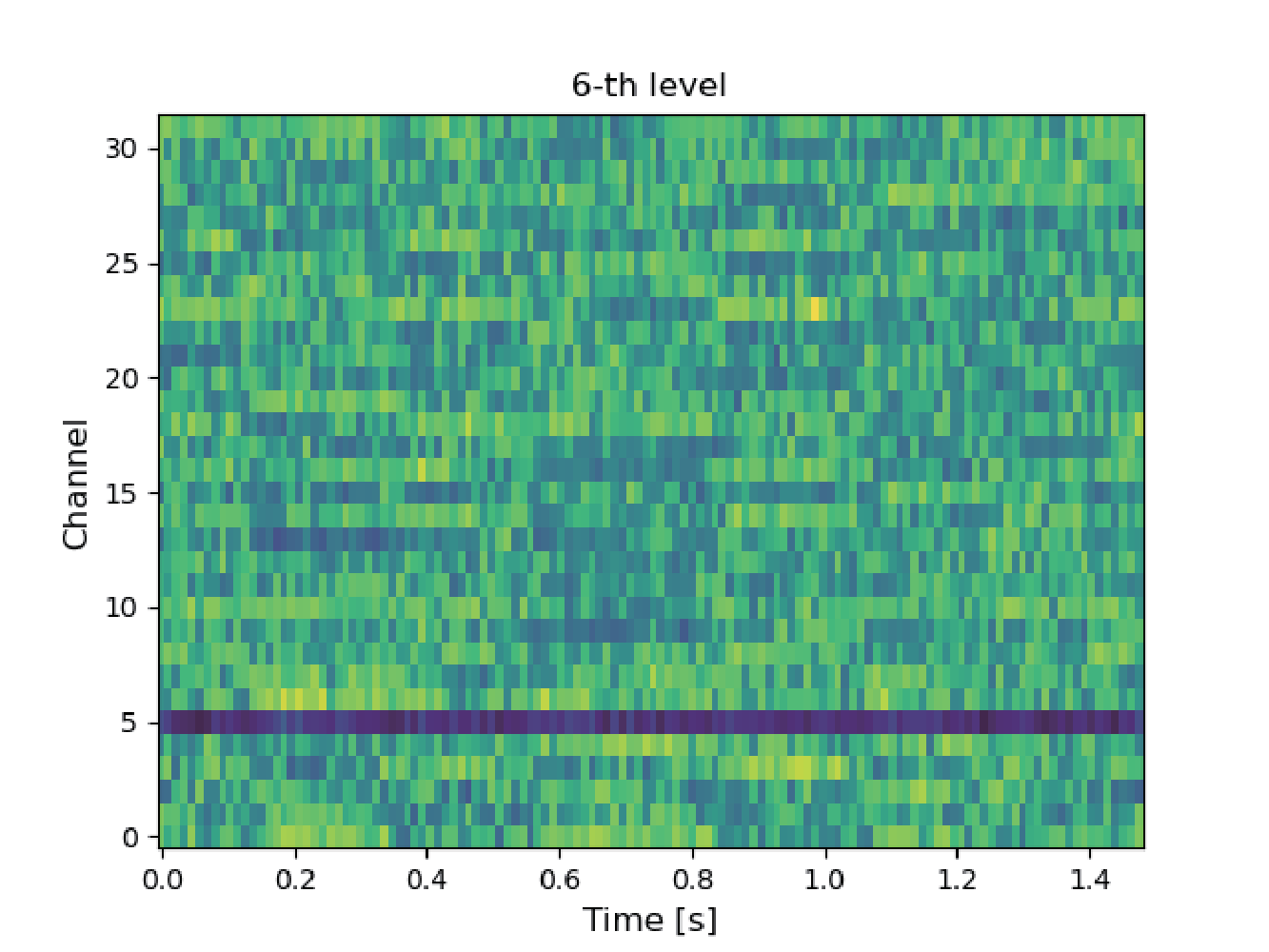}}
      \centerline{\includegraphics[width=.98\linewidth]{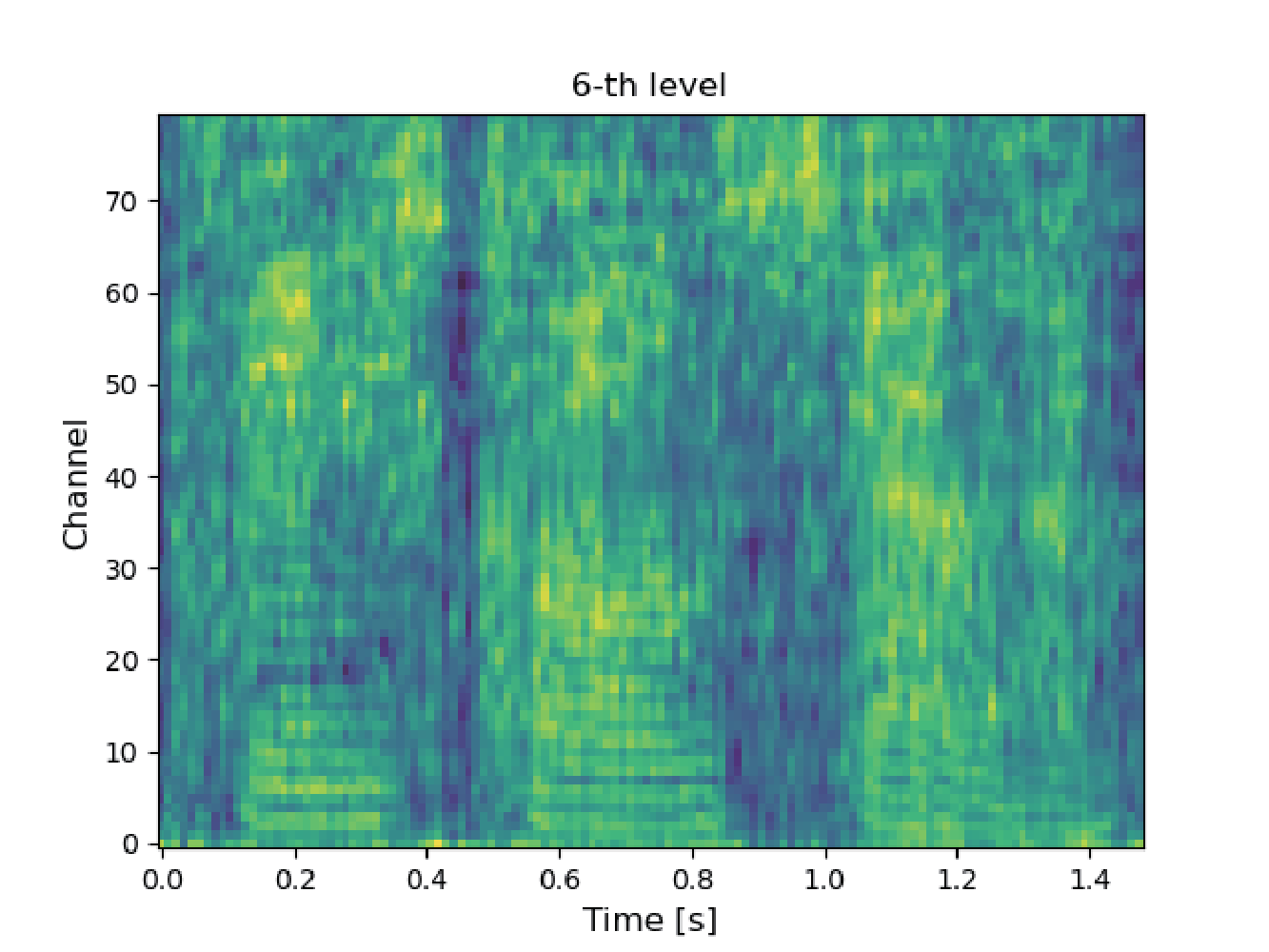}}
      \end{minipage}
\begin{minipage}[t!]{.245\linewidth}
        \centerline{\includegraphics[width=.98\linewidth]{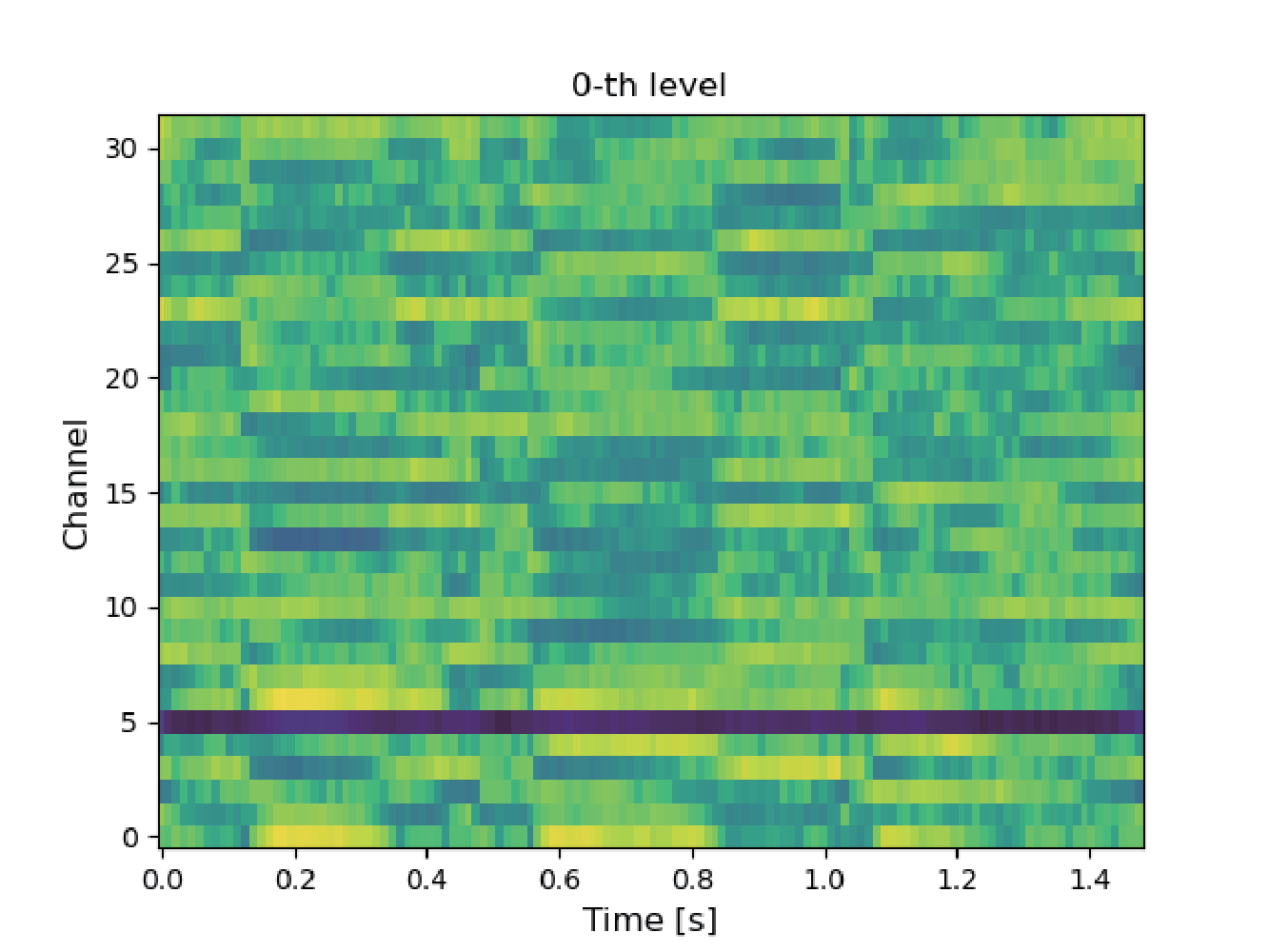}}
        \centerline{\includegraphics[width=.98\linewidth]{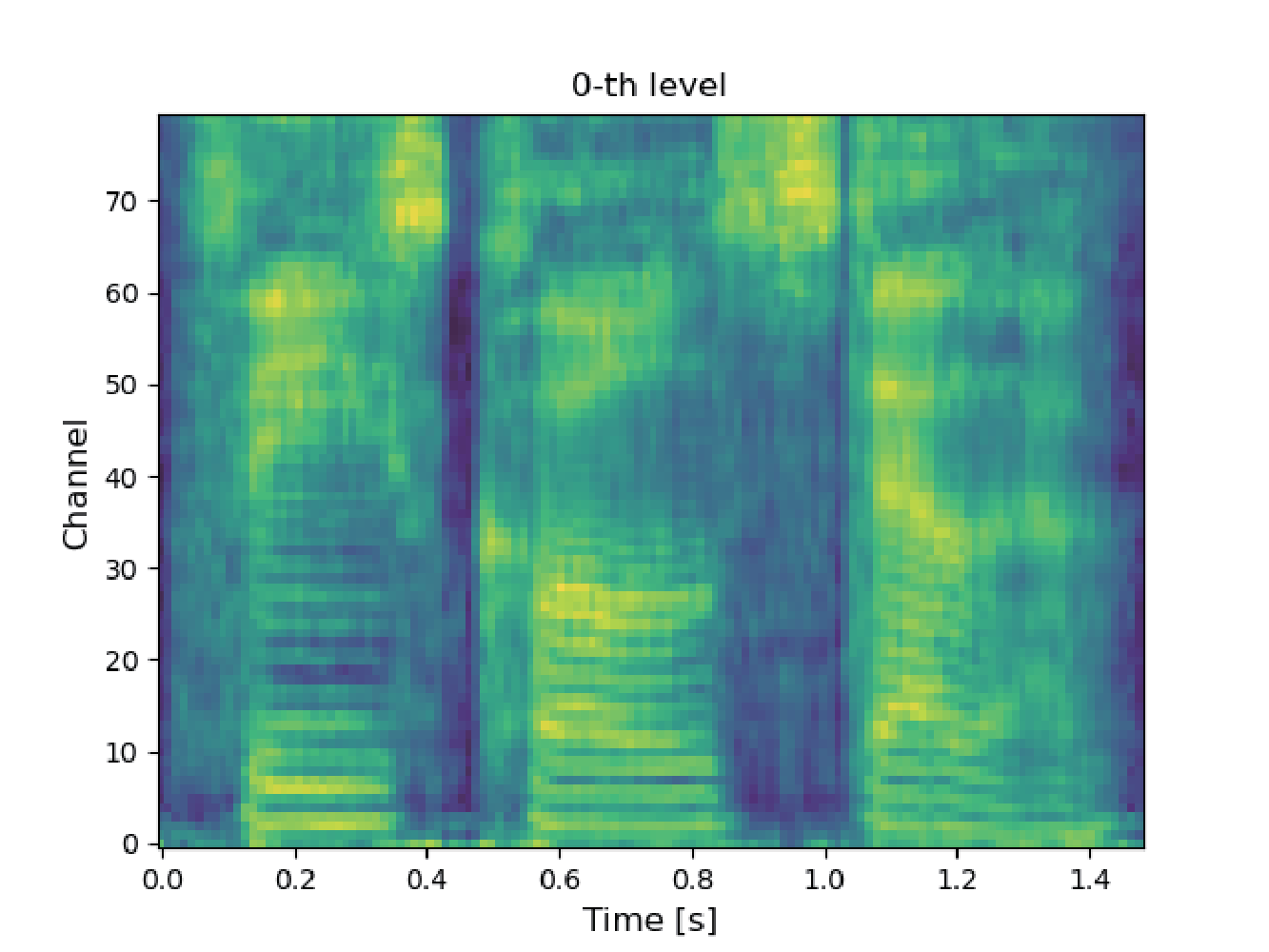}}
        \end{minipage}
 \vspace{-0ex}
  \centering
  \caption{Bottleneck feature sequence and mel-spectrogram updates (upper and lower rows, respectively) at steps 18, 12, 6, and 0 during LatentVoiceGrad-DPM's VC process, using speaker p238 (female)'s utterance as input and speaker p241 (male)'s utterance as the reference speech (from left to right).}
\label{fig:latentvoicegrad-dmp_vc-process}
\end{figure*}

\begin{figure*}[t!]
\centering
\begin{minipage}[t!]{.245\linewidth}
  \centerline{\includegraphics[width=.98\linewidth]{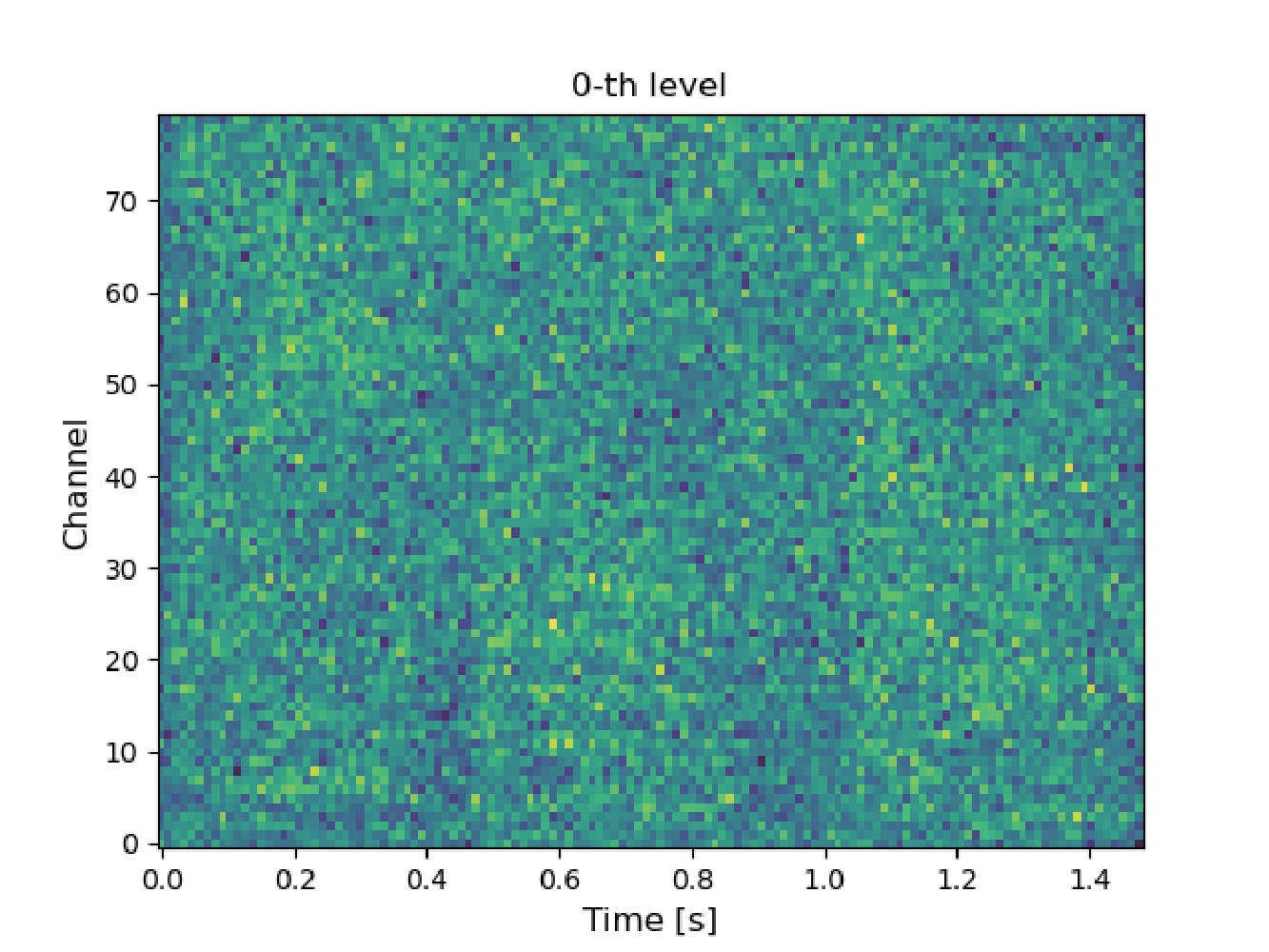}}
\end{minipage}
\begin{minipage}[t!]{.245\linewidth}
    \centerline{\includegraphics[width=.98\linewidth]{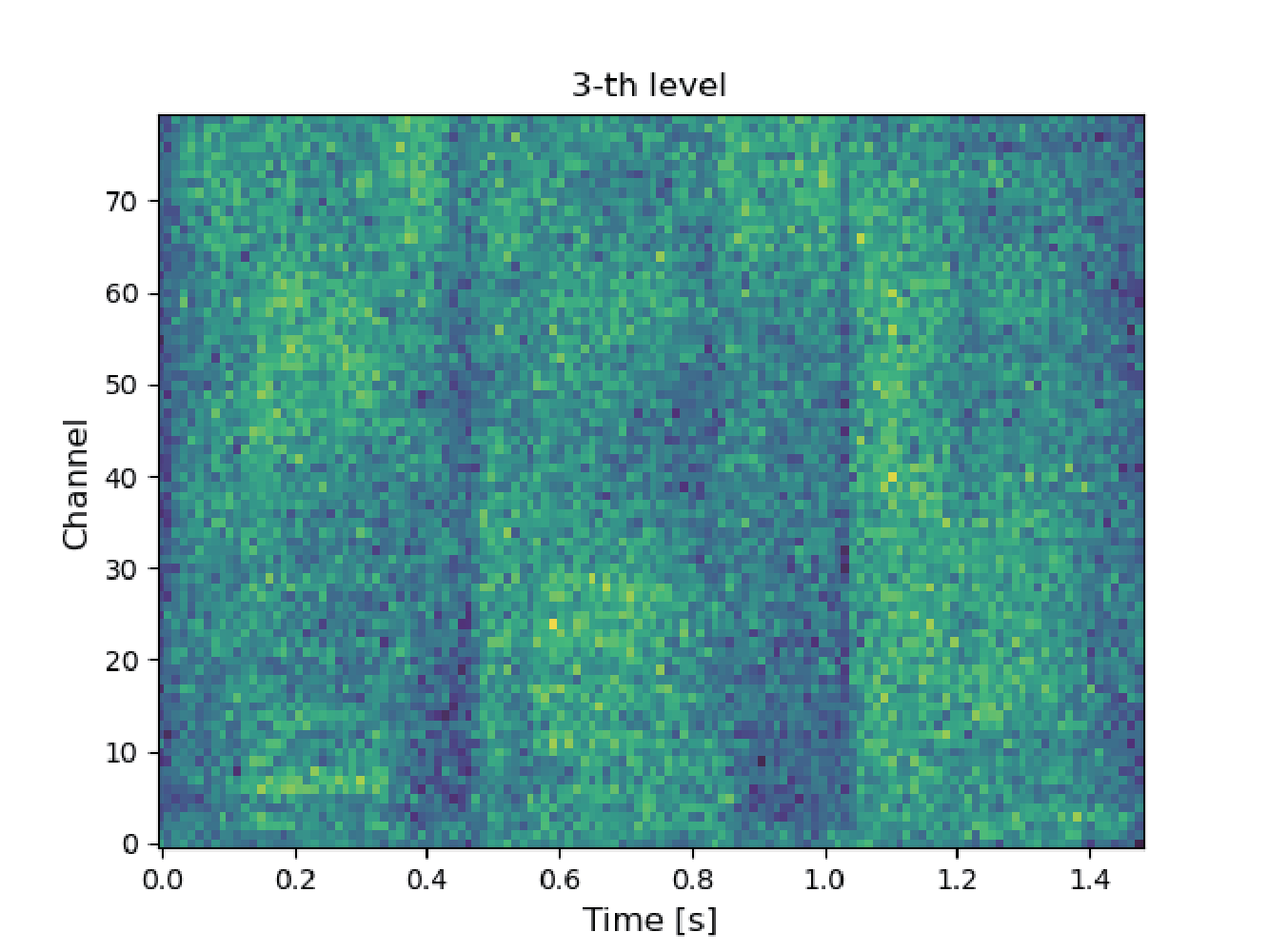}}
    \end{minipage}
\begin{minipage}[t!]{.245\linewidth}
      \centerline{\includegraphics[width=.98\linewidth]{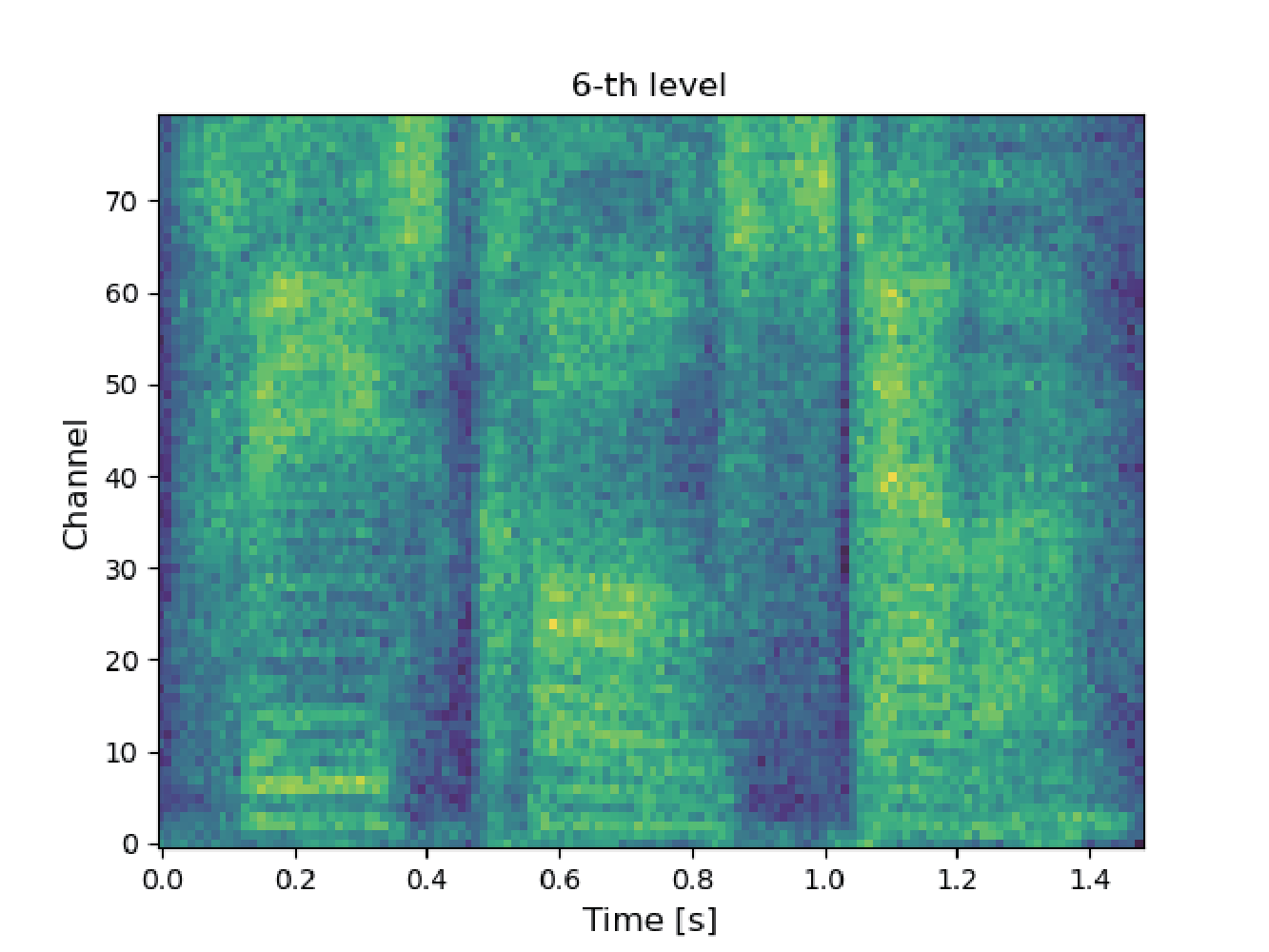}}
      \end{minipage}
\begin{minipage}[t!]{.245\linewidth}
        \centerline{\includegraphics[width=.98\linewidth]{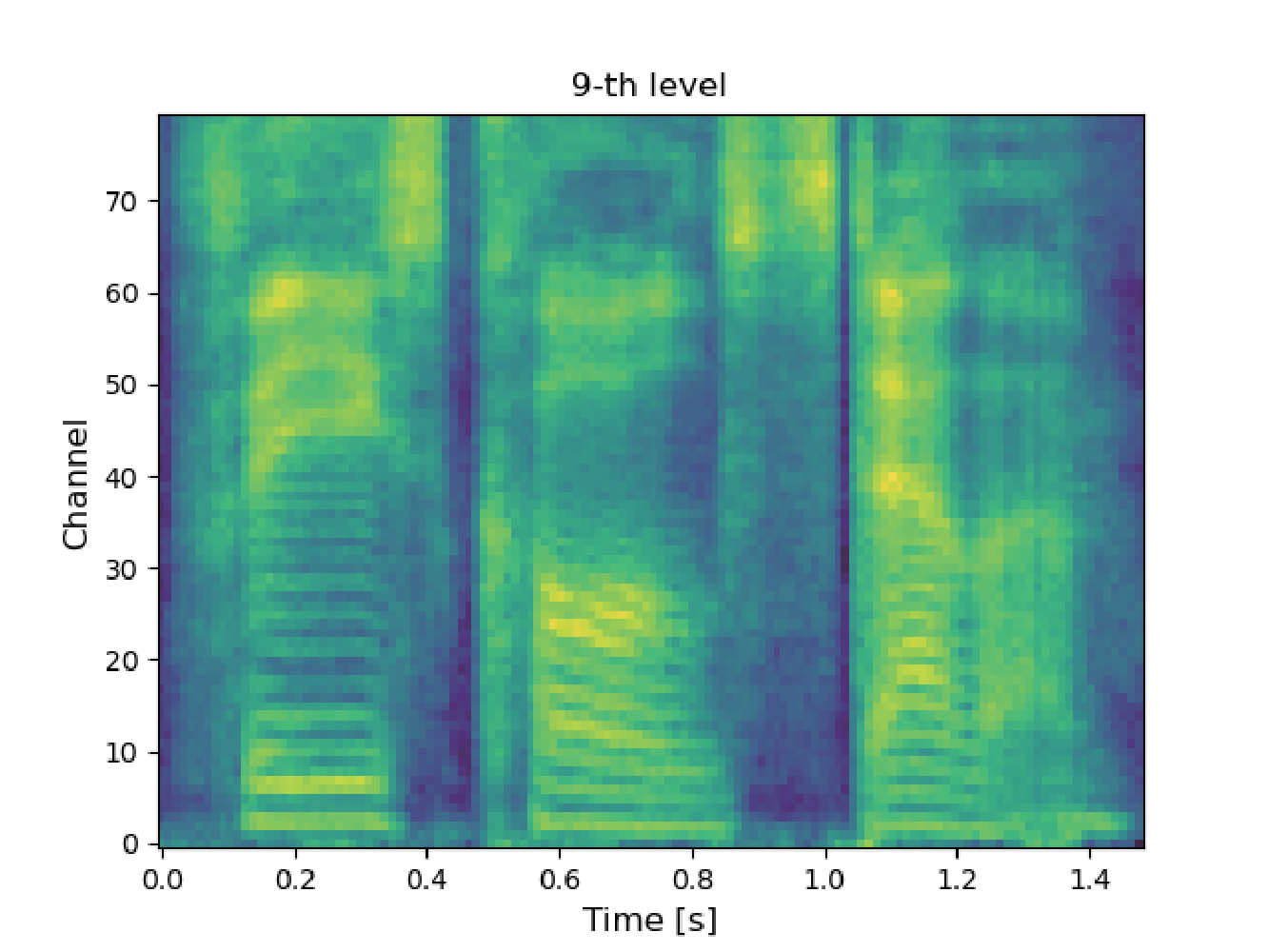}}
        \end{minipage}
 \vspace{-0ex}
  \centering
  \caption{Mel-spectrogram updates at steps 0, 3, 6, and 9 during VoiceGrad-FM's VC process, using speaker p238 (female)'s utterance as input and speaker p241 (male)'s utterance as the reference speech (from left to right).}
\label{fig:voicegrad-fm_vc-process}
\end{figure*}

\begin{figure*}[t!]
\centering
\begin{minipage}[t!]{.245\linewidth}
  \centerline{\includegraphics[width=.98\linewidth]{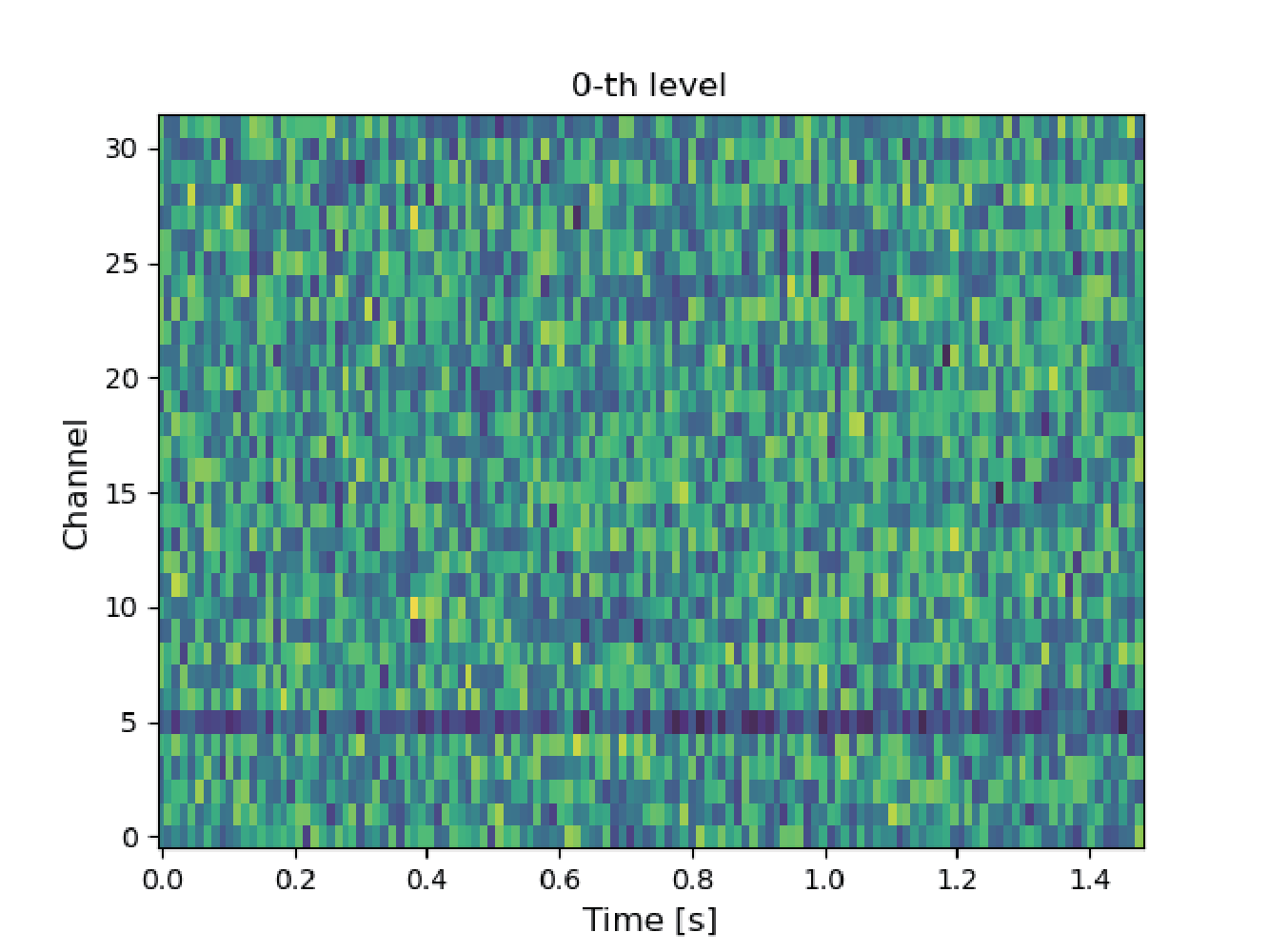}}
  \centerline{\includegraphics[width=.98\linewidth]{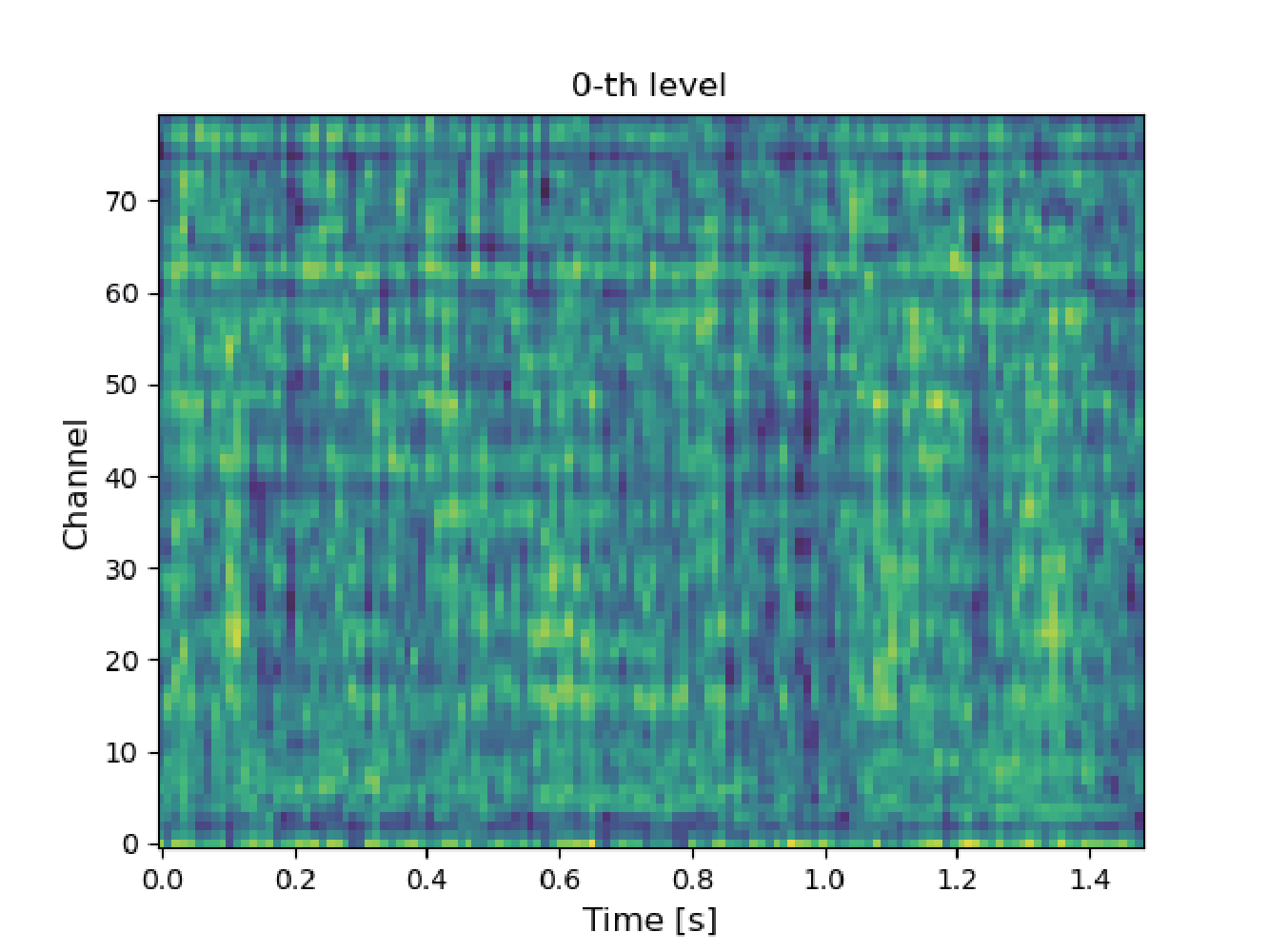}}
\end{minipage}
\begin{minipage}[t!]{.245\linewidth}
    \centerline{\includegraphics[width=.98\linewidth]{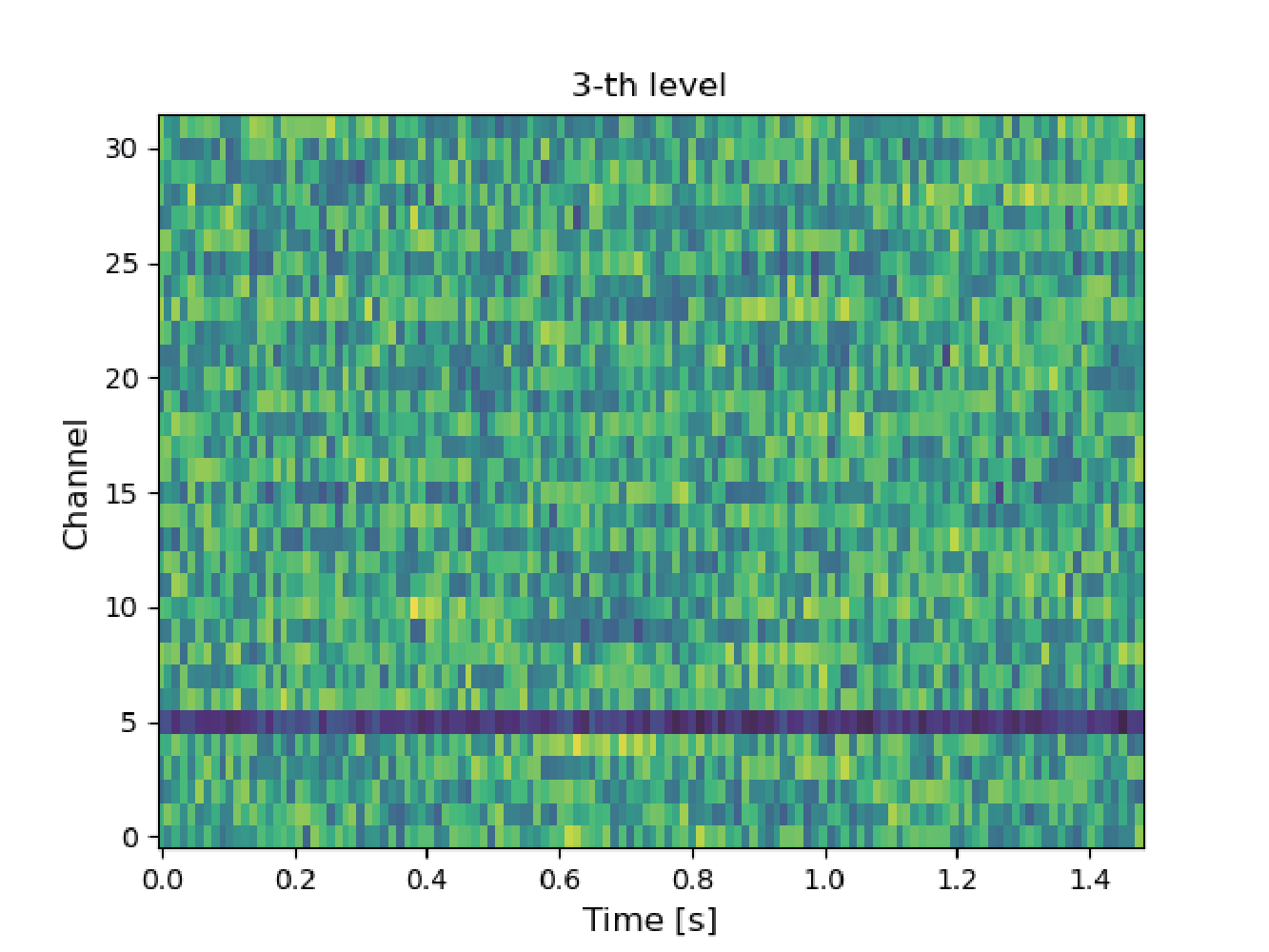}}
    \centerline{\includegraphics[width=.98\linewidth]{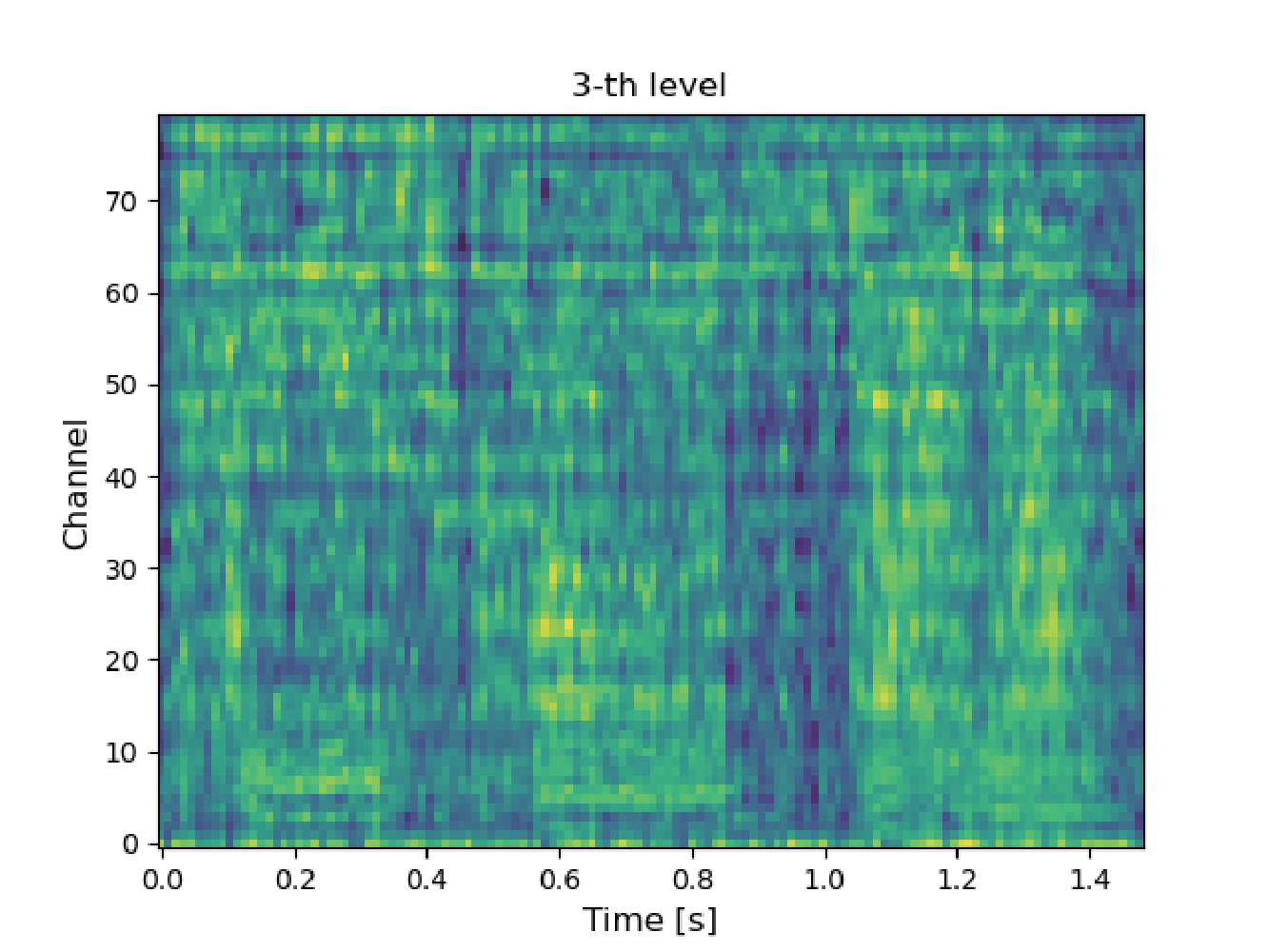}}
    \end{minipage}
\begin{minipage}[t!]{.245\linewidth}
      \centerline{\includegraphics[width=.98\linewidth]{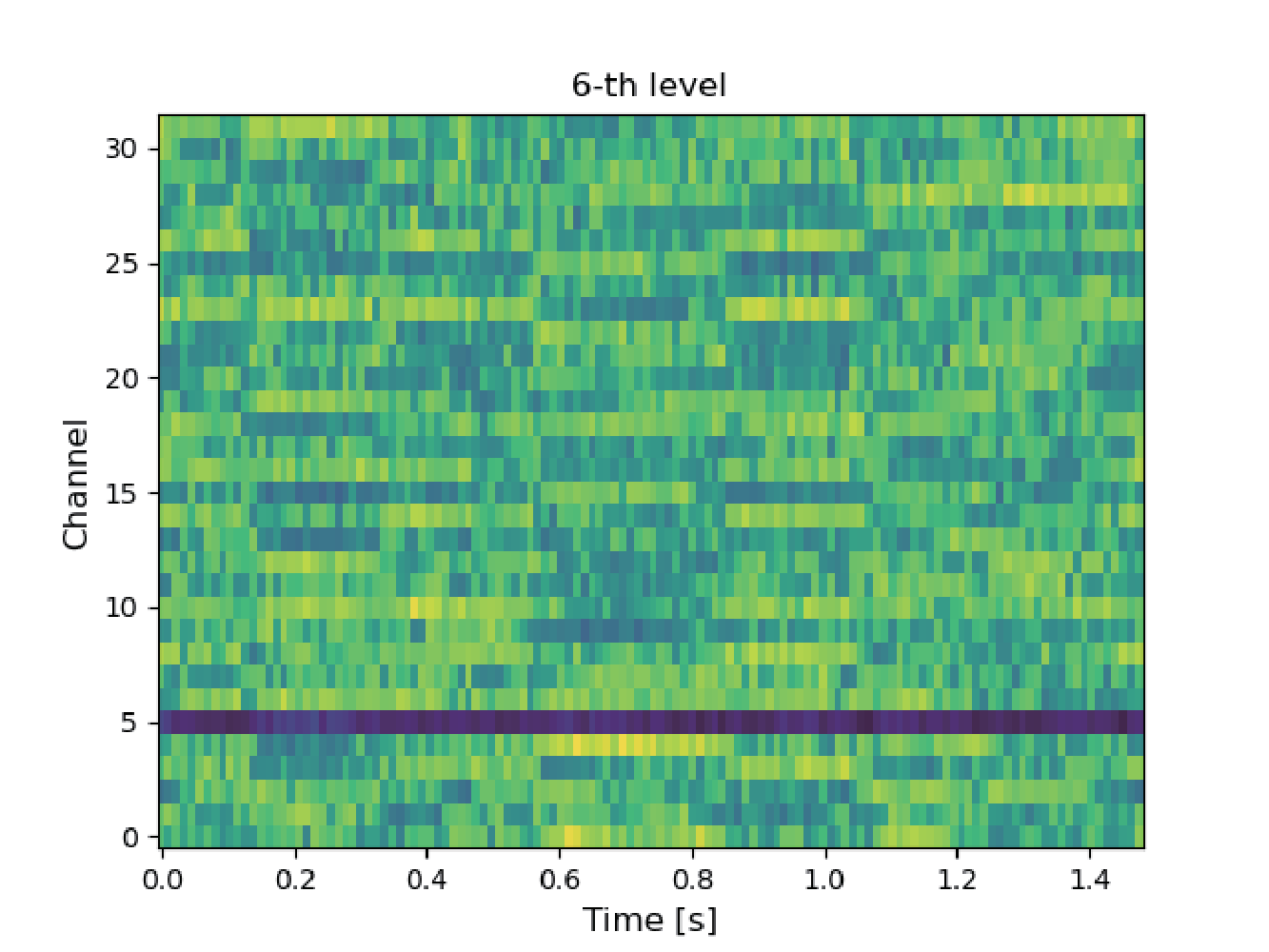}}
      \centerline{\includegraphics[width=.98\linewidth]{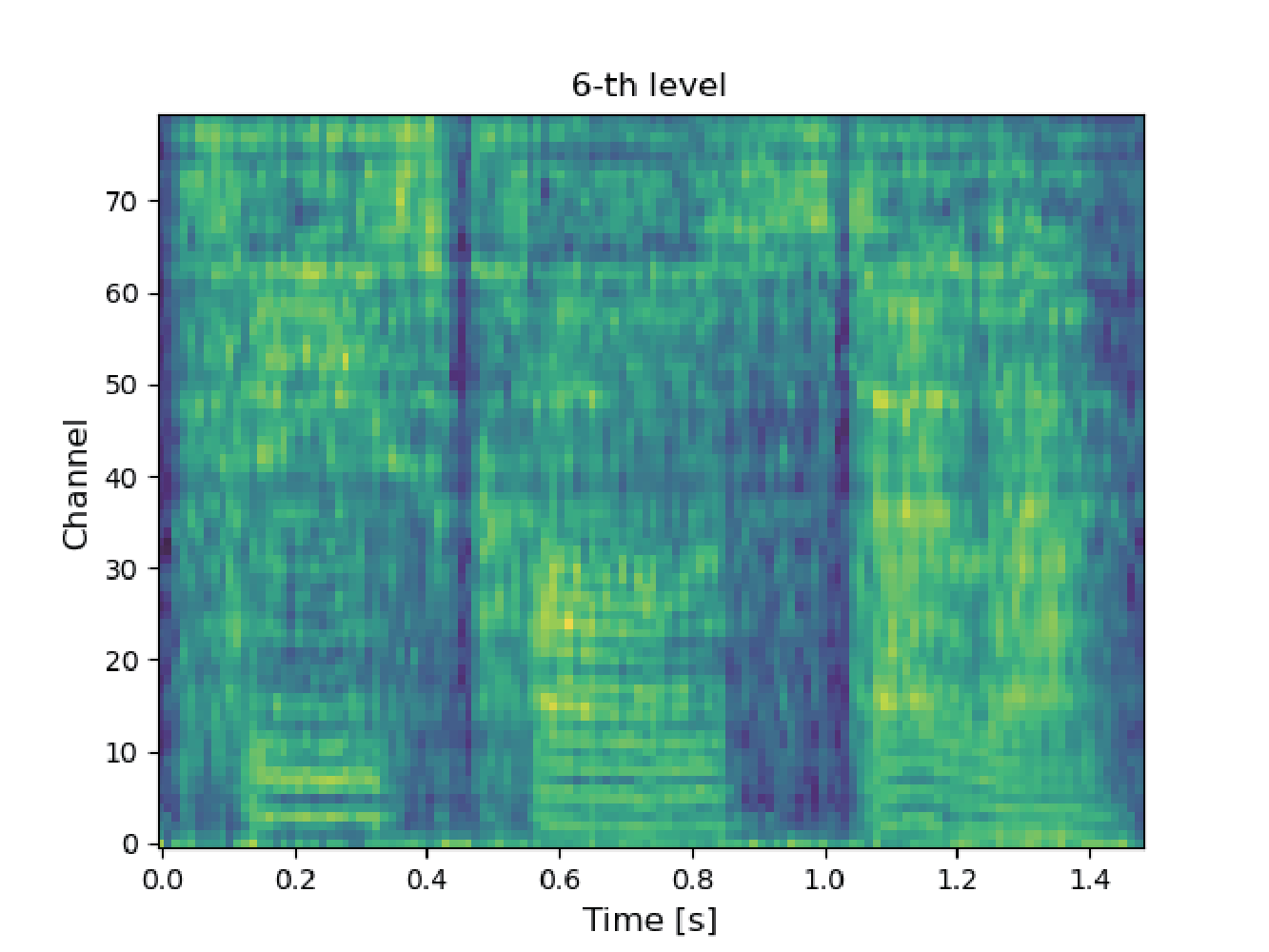}}
      \end{minipage}
\begin{minipage}[t!]{.245\linewidth}
        \centerline{\includegraphics[width=.98\linewidth]{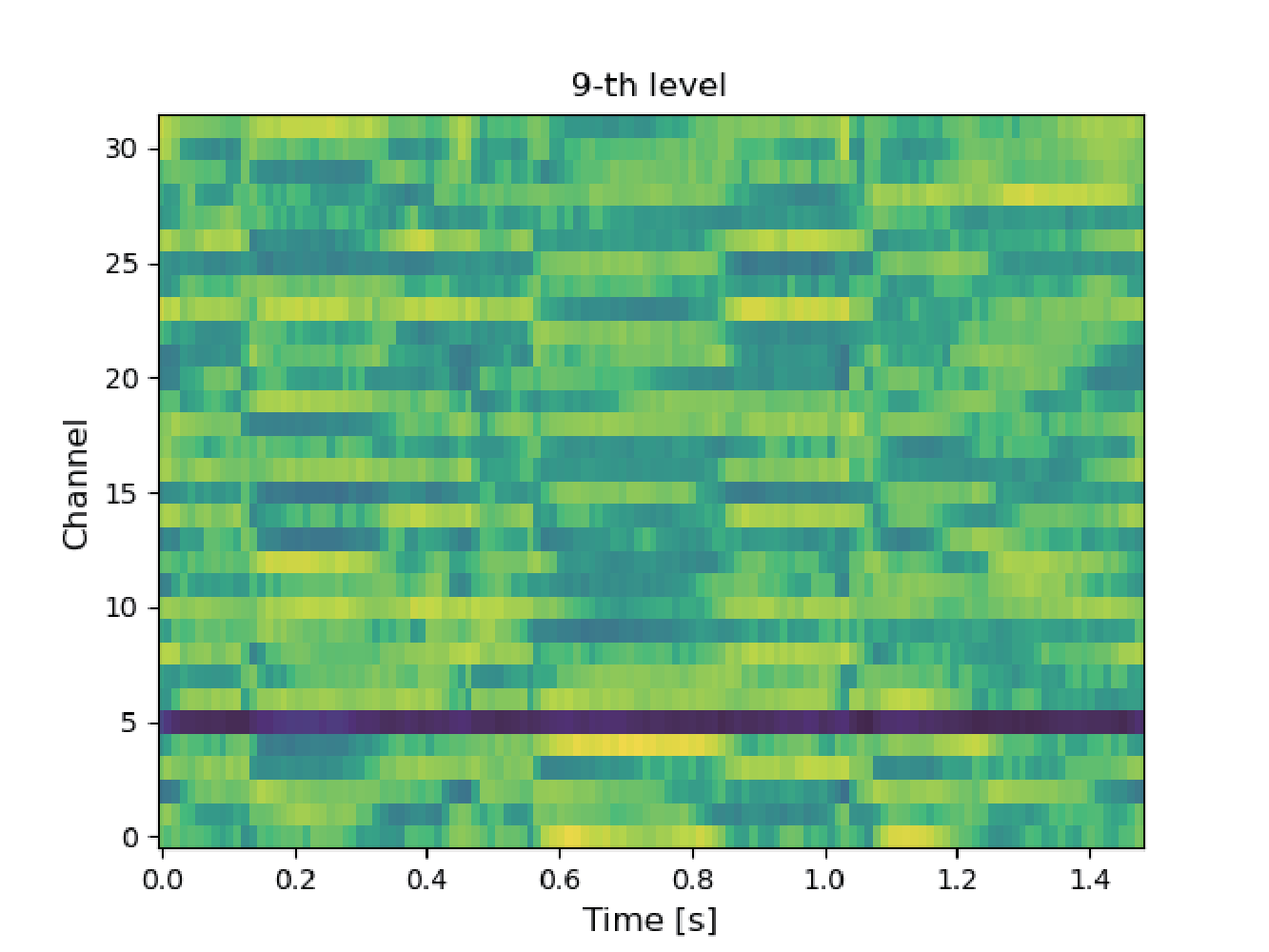}}
        \centerline{\includegraphics[width=.98\linewidth]{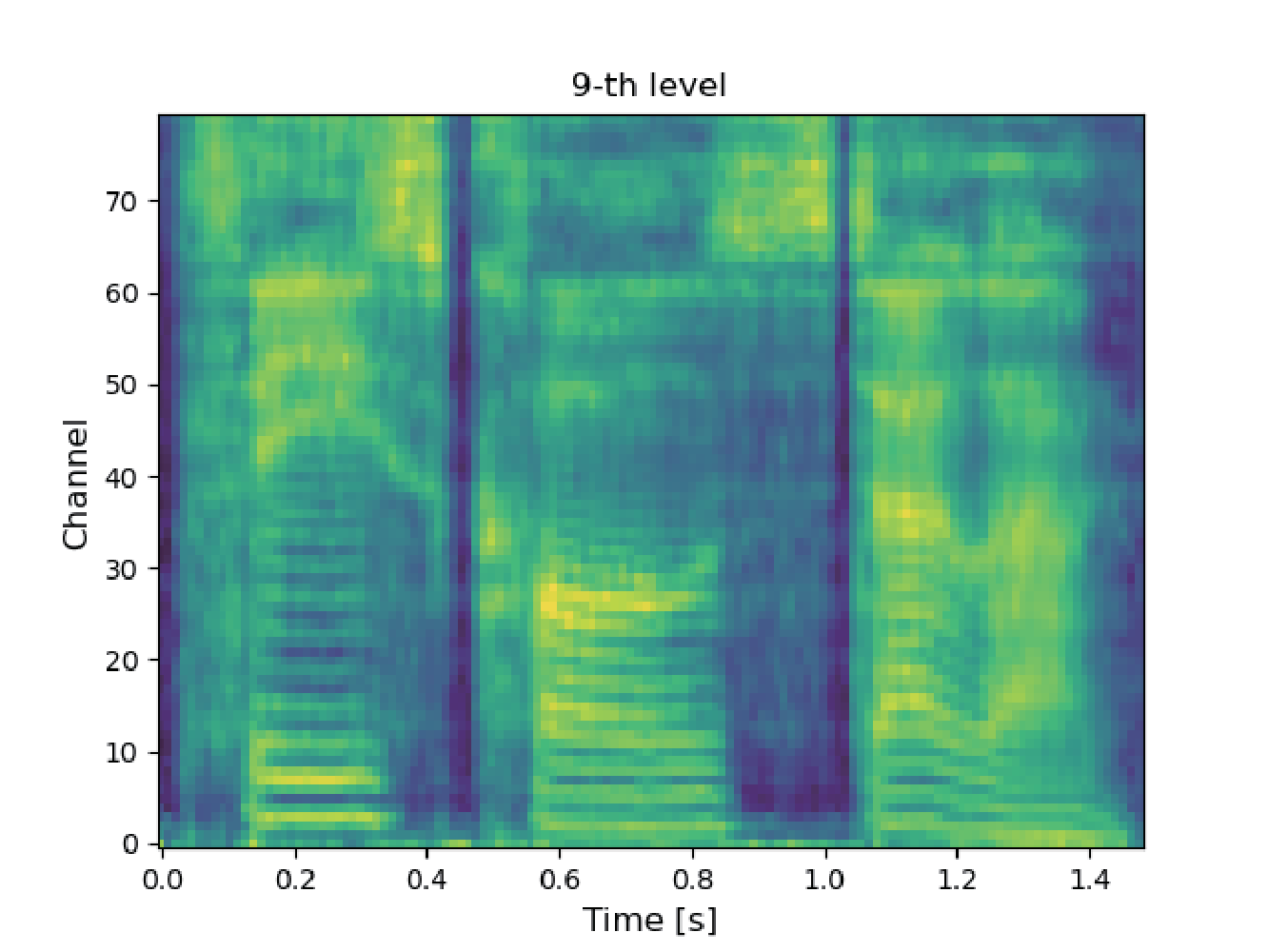}}
        \end{minipage}
 \vspace{-0ex}
  \centering
  \caption{Bottleneck feature sequence and mel-spectrogram updates (upper and lower rows, respectively) at steps 0, 3, 6, and 9 during LatentVoiceGrad-FM's VC process, using speaker p238 (female)'s utterance as input and speaker p241 (male)'s utterance as the reference speech (from left to right).}
\label{fig:latentvoicegrad-fm_vc-process}
\end{figure*}

\section{Conclusions}

In this paper, we explored two ideas for improving the performance and conversion speed of VoiceGrad, a VC method that frames mel-spectrogram conversion as the reverse diffusion process of a DPM. The first idea is to treat the latent representation of mel-spectrograms (bottleneck features) obtained from a pretrained speaker-independent autoencoder as the data to be converted. This autoencoder was pretrained with an adversarial loss that measures how realistic the waveform generated from the decoder and vocoder is, along with a regular reconstruction loss. The second idea is to replace the DPM with an FM model as the underlying generative model and undertake an ODE solver using a vector field network to perform VC.
Based on the experimental results, the following findings were confirmed: (1) incorporating adversarial loss during autoencoder training was beneficial; (2) for the FM version, mixing the initial point with Gaussian noise in a ratio of approximately 0.7:0.3 proved effective, and setting the number of iterations to around 10 achieved a good balance between audio quality and conversion performance; and (3) incorporating one or both of the ideas from the latent and FM versions resulted in higher performance than the original VoiceGrad and the baseline method (Diff-VC).

In our recent work, we proposed a method in which VoiceGrad serves as the teacher model, and its knowledge is distilled into a student model designed to perform mel-spectrogram conversion in a one-step process, achieving performance comparable to VoiceGrad with only one inference step \cite{Kaneko2024}. Given that the LatentVoiceGrad proposed in this paper surpasses VoiceGrad in both conversion quality and speed, it is expected that using LatentVoiceGrad as the teacher model could lead to the development of a one-step converter with even better performance in terms of both quality and training efficiency.

\section*{Acknowledgment}

This work was supported by JST CREST Grant Number JPMJCR19A3, Japan.

\ifCLASSOPTIONcaptionsoff
  \newpage
\fi



\bibliographystyle{IEEEtran}
\bibliography{Kameoka2024IEEETrans_LatentVoiceGrad}
%

\end{document}